\documentclass[conference,compsoc]{IEEEtran}
\IEEEoverridecommandlockouts

\usepackage{cite}
\usepackage{amsmath,amssymb,amsfonts}
\usepackage{algorithmic}
\usepackage{graphicx}
\usepackage{textcomp}
\usepackage{xcolor}
\usepackage{footmisc}
\usepackage{multirow}
\usepackage{array}
\usepackage{paralist}
\usepackage{url}
\usepackage{marvosym}

\newcommand{\skewsym}[1]{[#1]_{\times}}
\newcommand{\VecStyle}[1]{\mathbf{#1}}
\newcommand{\ScalarStyle}[1]{\mathit{#1}}
\newcommand{\Estimate}[1]{\Hat{#1}}
\newcommand{\Measurement}[1]{\Tilde{#1}}

\newcommand{\ModelState}{\VecStyle{x}}
\newcommand{\ModelInput}{\VecStyle{u}}
\newcommand{\ModelParams}{\mathcal{P}}

\newcommand{\bodyframe}{}

\newcommand{\deltatime}{\Delta {t}}

\newcommand{\unitvec}{\VecStyle{e_3}}

\newcommand{\pned}{\VecStyle{p}}

\newcommand{\vned}{\VecStyle{v}}

\newcommand{\Aned}{\mathbf{a}}

\newcommand{\dcm}{\mathbf{R}}
\newcommand{\quaternion}{\mathbf{q}}

\newcommand{\roll}{\phi}
\newcommand{\pitch}{\theta}
\newcommand{\yaw}{\psi}

\newcommand{\angrate}{\omega}
\newcommand{\AngRateVec}{\VecStyle{\angrate}}
\newcommand{\rollrate}{\angrate_\roll}
\newcommand{\pitchrate}{\angrate_\pitch}
\newcommand{\yawrate}{\angrate_\yaw}

\newcommand{\angacc}{\dot{\angrate}}
\newcommand{\AngAccVec}{\dot{\VecStyle{\angrate}}}
\newcommand{\rollacc}{\angacc_\roll}
\newcommand{\pitchacc}{\angacc_\pitch}
\newcommand{\yawacc}{\angacc_\yaw}

\newcommand{\CtrlTorque}{\VecStyle{m}^{\bodyframe}_{c}}

\newcommand{\CtrlAccel}{\VecStyle{a}^{\bodyframe}_{c}}

\newcommand{\DragAccel}{\VecStyle{a}^{\bodyframe}_{d}}

\newcommand{\InertiaMatrix}{\mathbf{I}}

\newcommand{\ThrustCoeff}{C_T}
\newcommand{\TorqueCoeff}{C_Q}
\newcommand{\TorqueRotorCoeff}{C_{Q, r}}
\newcommand{\ArmLength}{l}
\newcommand{\MotorRotRate}{\varpi}
\newcommand{\TimeConstant}{t_\textit{DC}}
\newcommand{\ThrustEstimate}{\Estimate{\ScalarStyle{T}}}
\newcommand{\ThrustSp}{\Bar{\ScalarStyle{T}}}
\newcommand{\ThrustSpZero}{\Bar{\ScalarStyle{T}}_\textit{Min}}
\newcommand{\ThrustSpRange}{\Bar{\ScalarStyle{T}}_\textit{Range}}

\newcommand{\VoltageLoad}{V_\textit{load}}
\newcommand{\CurrentLoad}{i_\textit{load}}

\newcommand{\VoltageRef}{V_\textit{ref}}
\newcommand{\ResistanceInternal}{R_\textit{int}}

\newcommand{\VehicleMass}{\mathcal{M}}
\newcommand{\Gee}{g}

\newcommand{\Airspeed}{\VecStyle{v}^{\bodyframe}_r}
\newcommand{\Windspeed}{\VecStyle{\vned}_w}
\newcommand{\EstimateWindSpeed}{\VecStyle{\Estimate{\vned}}_w}

\newcommand{\MCoef}{C_m}
\newcommand{\BCoefVec}{\VecStyle{C}_{bxy}}
\newcommand{\BCoefX}{C_{bx}}
\newcommand{\BCoefY}{C_{by}}

\newcommand{\TBuffer}{T_\textit{buf}}

\newcommand{\OvertAttack}{\text{OA}}
\newcommand{\StealthyAttack}{\text{SA}}
\newcommand{\MSAttack}[1]{\text{MA}_{\text{#1}}}
\newcommand{\GpsAttack}[1]{#1_\text{GPS}}
\newcommand{\GpsJointAttack}[1]{#1_\text{GPS-PV}}
\newcommand{\GyroAttack}[2]{\GyroAttackWithModel{#1}{#2}{Gyro}}
\newcommand{\GyroAttackWithModel}[3]{#1_\text{#3}^\text{#2/3}}
\newcommand{\AccelAttack}[2]{\GyroAttackWithModel{#1}{#2}{Accel}}
\newcommand{\BaroAttack}[2]{#1_\text{Baro}^\text{#2/2}}
\newcommand{\MagAttack}[2]{#1_\text{Mag}^\text{#2/2}}

\newcommand{\OAGps}{\GpsAttack{\OvertAttack}}
\newcommand{\OAGpsJoint}{\GpsJointAttack{\OvertAttack}}
\newcommand{\OATwoGyro}{\GyroAttack{\OvertAttack}{2}}
\newcommand{\OAThreeGyro}{\GyroAttack{\OvertAttack}{3}}
\newcommand{\OAGyroModel}[2]{\GyroAttackWithModel{\OvertAttack}{#1}{#2}}
\newcommand{\OAThreeAccel}{\AccelAttack{\OvertAttack}{3}}
\newcommand{\OABaro}{\BaroAttack{\OvertAttack}{2}}
\newcommand{\OAMag}{\MagAttack{\OvertAttack}{2}}
\newcommand{\SAGps}{\GpsAttack{\StealthyAttack}}
\newcommand{\SAGyro}{\GyroAttack{\StealthyAttack}{3}}

\newcommand{\sysname}{VIMU}
\newcommand{\sysnamefull}{Virtual IMU}
\newcommand{\se}{\sysname{}-SE}
\newcommand{\ad}{\sysname{}-AD}
\newcommand{\recovery}{\sysname{}-RM}
\newcommand{\physical}{\sysname{}-PM}

\newcommand{\ControlInvariant}{CI}
\newcommand{\SoftwareSensor}{SRR}
\newcommand{\savior}{SAVIOR}
\newcommand{\saviorbuffer}{\savior{}-Buffer}
\newcommand{\vimucusum}{\sysname{}-CS}
\newcommand{\vimuNoBuffer}{\sysname{}-NoBuffer}

\newcommand{\revise}[1]{\textcolor{black}{#1}}

\def\BibTeX{{\rm B\kern-.05em{\sc i\kern-.025em b}\kern-.08em
    T\kern-.1667em\lower.7ex\hbox{E}\kern-.125emX}}
\begin{document}

\title{\sysname{}: Effective Physics-based Realtime Detection and Recovery\\ against Stealthy Attacks on UAVs
\thanks{The post-conference authors’ version fixed several figure display issues (Figures 7, 8, 9, 16, 18, 19, 21) in the production of the conference-proceeding version.}
}

\author{
\IEEEauthorblockN{\normalsize
Yunbo~Wang\IEEEauthorrefmark{1},
Cong~Sun\IEEEauthorrefmark{1}\textsuperscript{\Letter}\thanks{\Letter~Corresponding author.},
Qiaosen~Liu\IEEEauthorrefmark{1},
Bingnan~Su\IEEEauthorrefmark{1},
Zongxu~Zhang\IEEEauthorrefmark{1},
Michael~Norris\IEEEauthorrefmark{2},
Gang~Tan\IEEEauthorrefmark{2},
Jianfeng~Ma\IEEEauthorrefmark{1}}
\IEEEauthorblockA{\normalsize \IEEEauthorrefmark{1} School of Cyber Engineering, Xidian University, China}
\IEEEauthorblockA{\normalsize \IEEEauthorrefmark{2} The Pennsylvania State University, University Park, PA, USA}
\normalsize Email: robertwang@stu.xidian.edu.cn, suncong@xidian.edu.cn, \{man5336, gtan\}@psu.edu
}

\maketitle

\begin{abstract}
Sensor attacks on robotic vehicles have become pervasive and manipulative. Their latest advancements exploit sensor and detector characteristics to bypass detection. Recent security efforts have leveraged the physics-based model to detect or mitigate sensor attacks.
However, these approaches are only resilient to a few sensor attacks and still need improvement in detection effectiveness.
We present \sysname{}, an efficient sensor attack detection and resilience system for unmanned aerial vehicles.
We propose a detection algorithm, CS-EMA, that leverages low-pass filtering to identify stealthy gyroscope attacks while achieving an overall effective sensor attack detection.
We develop a fine-grained nonlinear physical model with precise aerodynamic and propulsion wrench modeling.
We also augment the state estimation with a FIFO buffer safeguard to mitigate the impact of high-rate IMU attacks.
The proposed physical model and buffer safeguard provide an effective system state recovery toward maintaining flight stability.
We implement \sysname{} on PX4 autopilot.
The evaluation results demonstrate the effectiveness of \sysname{} in detecting and mitigating various realistic sensor attacks, especially stealthy attacks.
\end{abstract}


\section{Introduction}\label{sec:intro}

Airborne drones rely on sensor measurements for navigation and flight control. The sensors continually transduce raw physical signals into digital forms that can be interpreted and operated by the controller software~\cite{fu2018risks}. For instance, the autopilot controller uses gyroscope measurements to track changes in flight attitude.
Autopilots use simple validation (e.g., majority voting) by default to isolate the faulty sensor.
Although this proved efficient in handling hardware failures, this practice is being challenged by increasingly sophisticated sensor attacks.

Launching GPS spoofing attacks to take over robotic vehicles is straightforward~\cite{tippenhauer2011requirements, kerns2014unmanned},
and the evolved spoofing practices have been further facilitated by the widespread low-cost software-defined radio platforms~\cite{noh2019tractor, shen2020drift}.
Meanwhile,
another primary sensor attack category targets the inertial measurement unit (IMU). An IMU consists of an accelerometer, a gyroscope, and an optional magnetometer. The accelerometer measures changes in velocity, whereas the gyroscope and magnetometer are critical to attitude control. If they report erroneous measurements, the drone will immediately lose control and crash~\cite{son2015rocking, jang2023paralyzing}. Hence, effective attacks have been demonstrated by tampering with the accelerometer or gyroscope readings using sound waves at a known resonant frequency of the target sensor~\cite{son2015rocking, trippel2017walnut, tu2018injected} or remotely blocking the transmissions between IMU and flight controller through electromagnetic interference (EMI)~\cite{jang2023paralyzing}.

Conventional security mechanisms, e.g., data encryption, network protection, or software-oriented mitigation, are inadequate to protect robotic vehicles against sensor attacks since such attacks are conducted on an orthogonal attack surface~\cite{fu2018risks}.
People have used the physical invariants intrinsic in cyber-physical systems to detect or mitigate sensor attacks~\cite{10.1145/1952982.1952995, 10.1145/3243734.3243781, 10.1145/3264888.3264893, giraldo2018survey}.
This physics-based approach also received growing attention in robotic vehicles due to its explainability and low overhead~\cite{choi2018detecting, quinonez2020savior, choi2020software}. The physics-based approaches are categorized into \emph{physics-based attack detection} (PBAD) and \emph{physics-based attack resilience} (PBAR). Both rely on a physical model that makes iterative predictions on the expected system states. Generally, the PBAD approaches compare the predictions with sensor measurements to detect the sensor attacks. The PBAR approaches, on the other hand, strive to sustain system operation by using the physical model to recover or replace the effects of faulty sensors.

Pushing towards efficient PBAD and PBAR in robotic vehicles, recent efforts have used the linear state-space model to capture the physical invariants in the vehicles~\cite{choi2018detecting, choi2020software}.
Using a linear model to approximate the higher-order dynamic has been justified as analogous to the PID control~\cite{choi2020software}.
However, PID control relies on accurate feedback from the functional sensors to correct the approximation error.
In contrast, PBAD and PBAR are designed to reject the compromised sensors, and their physical model should continue the state prediction even without the correction from sensors.
Without such sensor-based correction, the model performance will deteriorate as the approximation error accumulates.
We expect a more accurate physical model, e.g., a well-fitted and high-order nonlinear model, to delay this process and take effect as the basis of sensor-attack resilience. Nonlinear physical invariant has been deployed in PBAD~\cite{quinonez2020savior} without considering recovery.
This nonlinear physical model does not account for the propulsion losses caused by the electrical and mechanical factors, which impede its usage in attack resilience.
Other approaches based on passive fault-tolerant control (FTC)~\cite{DBLP:conf/dsn/DashLCKP21, fei2020learn, tu2019flight} also apply nonlinear models to capture the high-order vehicle dynamic.
However, these solutions' high computational overhead and processing latency make mitigating high-rate IMU attacks harder~\cite{jang2023paralyzing}.

Besides the model's accuracy and efficiency, there are more security concerns in state-of-the-art approaches. First,
existing PBAD and PBAR approaches~\cite{quinonez2020savior, choi2018detecting, choi2020software} use statistic-based attack detection.
However, they are limited in differentiating the measurement noise of each sensor (e.g., gyroscope) from the detection statistics, leaving a loose alarm threshold for advanced adversaries to conduct stealthy attacks.
Second, high-rate IMU attacks can crash the drone within one sampling interval~\cite{jeong2023unrock, jang2023paralyzing}. Existing PBAD and PBAR approaches cannot identify such attacks before the faulty data enters the estimator and control loop.
Third, passive FTC approaches~\cite{fei2020learn, DBLP:conf/dsn/DashLCKP21} accept all sensor measurements without validation, allowing the adversary to manipulate the flight control freely through vulnerable sensor data.
Fourth, the state-of-the-art PBAR approach~\cite{choi2020software} requires a functional accelerometer for recovery when all gyroscopes are compromised. As both sensors belong to IMU, such recovery dependency is exposed to attacks that corrupt all transmissions between IMUs and autopilot~\cite{jang2023paralyzing}.

Regarding the above limitations, we emphasize four criteria for an effective PBAR: 1) an accurate physical model that supports long recovery duration, 2) an effective detection algorithm with tightened thresholds to restrain stealthy sensor attacks, 3) a delay mechanism avoiding high-rate attack data reaching the state estimates before being detected, and 4) an efficient implementation meeting the timing and resources of real hardware. In this work, we present \sysnamefull{} (\sysname{}), a new PBAR framework to address sensor attacks, esp. stealthy attacks, against aerial vehicles.
We tighten alarm thresholds with a new detection algorithm, CS-EMA.
CS-EMA extends \emph{Cumulative Sum} (CUSUM) with a residual-capped EMA detector, which also works as a low-pass filter to distinguish the persistent deviation from high-frequency measurement noises.
Then, we use a nonlinear physical model that includes precise submodels on aerodynamic drag and propulsion wrench to achieve a recovery performance better than the state-of-the-art PBAR~\cite{choi2020software} and the nonlinear model of~\cite{quinonez2020savior}.
To address the high-rate IMU attacks, we design a FIFO buffer safeguard that prevents faulty data from reaching the state estimates immediately.
We summarize our contributions as follows:
\begin{compactenum}[1.]
  \item We found that specific measurement noise, e.g., the gyroscope's white noise, can be filtered out to improve detection effectiveness. Our detection algorithm, CS-EMA, outperforms the detection algorithms of the state-of-the-art PBAD~\cite{quinonez2020savior, choi2018detecting} and PBAR~\cite{choi2020software} in effectiveness against the overt and stealthy gyroscope attacks of \cite{quinonez2020savior, gao2023exploring, tu2018injected}. \revise{CS-EMA detector identifies more stealthy attacks than the detectors of~\cite{quinonez2020savior, choi2018detecting, choi2020software}.}

  \item \revise{We propose the buffer safeguard to protect the integrity of UAV's reference states from instant attack injection. With the fine-grained physical model} and buffer safeguard, VIMU surpasses the existing PBAR~\cite{choi2020software} and the nonlinear physical model of~\cite{quinonez2020savior} in recovery duration. Even when all IMUs are compromised, \sysname{} can sustain flight stability longer without relying on the accelerometer or external system, as opposed to~\cite{choi2020software}.

  \item We implement \sysname{} in PX4 autopilot. We demonstrate that our implementation meets the timing and resources on real hardware (CUAV v5+). Our implementation is available at \url{https://github.com/wangwwno1/Project-VIMU}.

\end{compactenum}

\section{Adversary Model}\label{subsec:threat model}

We target a similar adversary model to \cite{choi2020software, quinonez2020savior}.
The adversary can inject false signals into multiple sensors and obtain the following knowledge to carry out the attack:
\begin{compactenum}[1.]
  \item Vehicle's hardware specifications and physical properties.
  \item The parameter settings of the autopilot program and its anomaly detector, if the anomaly detector is available.
  \item The maneuver commands from the autopilot or human operator.
\end{compactenum}
Our detection approach identifies sensor attacks based on the deviation of sensor measurements from the predicted state of the physical model, which requires an accurate system state estimation at the initialization of \sysname{}.
Since \sysname{} is designed to launch simultaneously with the autopilot controller, we assume the autopilot controller can acquire the drone's accurate initial system state, including position, velocity, and attitude.
We assume the actuators' correct functionality to ensure our physical model can predict the system states with the actuator commands from the autopilot controller. Thus, handling the faulty actuators is out of our scope.
The faulty actuators can be addressed by the fault-tolerant control approaches~\cite{zhang2020real, zhang2021real, fei2020learn, DBLP:conf/dsn/DashLCKP21},  but such approaches require further efforts to reduce computational delay and meet the real-time constraints.
Although our approach can tolerate considerable wind disturbances on GPS and gyroscope, detecting and mitigating more violent external disturbances, e.g., physical collisions or the proximity to obstacles, are out of our scope. These disturbances can be discriminated through existing solutions, e.g., contact detection~\cite{tomic2020simultaneous} or ground effect model~\cite{bangura2012nonlinear}.
Moreover, the attack surface of software and cyber attacks is orthogonal to the sensor attacks addressed in this work. Mitigating these threats is also beyond our scope.

\section{Design of \sysname{}}
\label{sec:system_design}

\begin{figure}[!tb]
    \centering
    \includegraphics[width=0.9\columnwidth]{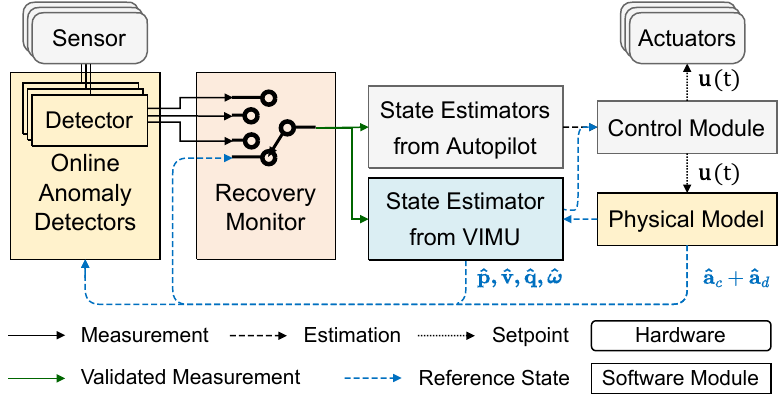}
\vspace{-1ex}
    \caption{High-level Overview of \sysname{} Modules}
    \label{fig:padr_workflow}
\vspace{-3ex}
\end{figure}

\sysname{} is a physics-based attack resilience solution.
It attains attack resistance by providing the autopilot with uninterrupted state estimates even when the sensors are subjected to attack. \sysname{} has four modules: \emph{anomaly detectors}, a \emph{recovery monitor}, a \emph{state estimator}, and a \emph{physical model}.
Fig.~\ref{fig:padr_workflow} shows the block diagram of \sysname{}.
The anomaly detectors (\ad{}) identify compromised sensors.
The recovery monitor (\recovery{}) selects dependable data sources based on the detection results.
The state estimator (\se{}) combines predictions from the nonlinear physical model (\physical{}) with available sensor measurements to provide a reliable \emph{estimate} of system states.
We denote the estimates from \se{} and \physical{} as \emph{reference states}, in contrast to the estimates provided by state estimators from the autopilot.
When the system is not under attack, \sysname{} ensures the estimates from the vanilla estimators converge to the reference states, as both estimated states are subject to the same physical laws.
\sysname{} carries out the following functionalities to protect flight safety:

\begin{compactenum}[1.]
\item \emph{Anomaly detection and recovery}.
        The first step towards attack resilience is to detect and isolate compromised sensors.
        The anomaly detector compares the reference states with sensor measurements to determine whether the sensor instance is compromised (Section~\ref{subsec:method_detection}).
        The recovery monitor enforces the lattice-based recovery policies, filters out compromised sensor data, and delivers the remaining measurements to all state estimators (Section~\ref{subsec:method_recovery}). As compromised sensors get isolated, the state estimators with available inputs, including \se{}, accurately estimate the system state, ensuring the safe control of autopilot.
\item \emph{Uninterrupted estimation for flight control}.
    Vanilla state estimators in autopilot rely on IMU measurements to update their estimates.
    This design fails when all IMU instances are compromised. Accepting false sensor data disturbs the estimated state. On the other hand, isolating the compromised sensor instances will interrupt the vanilla estimators updating estimates. In both cases, the flight controller no longer tracks the system state, leading to an instant control loss.
    The reference state serves as a backup of existing estimates (Section~\ref{subsec:method_recovery}).
    The physical model (Section~\ref{subsec:physical model}) takes the last setpoint of actuator output issued by the controller as the control input $\ModelInput(t)$ to update its reference states.
    \se{} takes these reference states and the measurements of uncompromised IMUs to update the estimates of \se{} (Section~\ref{subsec:method_ekf}).
    This update process is iterative and works even without an operational IMU.
    When all IMUs get compromised, the control module switches to the reference states for flight control, thus ensuring continuous operation under attack.
\end{compactenum}

\subsection{Nonlinear Physical Model}\label{subsec:physical model}

\begin{table}[!t]
\renewcommand{\arraystretch}{1.2}
  \caption{Physical Parameters of Nonlinear Physical Model}
  \label{tab:physical-parameters-definition}\scriptsize
\vspace{-1ex}
\resizebox{\columnwidth}{!}{
  \begin{tabular}{c l}
    \hline
    \textbf{Param} & \textbf{Definition} \\
    \hline
    $\VehicleMass$ & Mass of the drone \\
    $\ArmLength$ & Distance from the motor to the airframe CoG \\
    $\InertiaMatrix$ & 3$\times$3 symmetric matrix of airframe's moment of inertia. It comprises 3 \\
      & diagonal terms ($I_{xx},I_{yy},I_{zz}$) and 3 non-diagonal terms ($I_{xy},I_{xz},I_{yz}$) \\
    $\ThrustCoeff$ & thrust coefficient \\
    $\TorqueCoeff$ & torque coefficient \\
    $\TorqueRotorCoeff$ & coefficient of rotor's gyroscopic moment \\
    $\TimeConstant{}$ & time constant for the rotor speed to rise or fall \\
    $\VoltageRef$ & nominal voltage when the battery module has charged to full capacity \\
    $\ThrustSpZero$ & the minimum command to spin up the rotor \\
    $\ThrustSpRange$ & the range of the motor command \\
    $\ResistanceInternal$ & battery internal resistance \\
    $\MCoef$ & momentum drag coefficient \\
    $\BCoefVec$ & horizontal ballistic coefficient vector $[\BCoefX, \BCoefY, 0]^T$, s.t. $\BCoefX$ and $\BCoefY$ \\
    & are the forward and right direction of airframe, respectively \\
  \hline
\end{tabular}
}
\vspace{-4ex}
\end{table}

The drone's airframe type impacts the concrete design of the physical model. However, different physical models generally follow the same development process. This work uses a quadcopter as the demonstration platform.
Without loss of generality, we follow the common assumptions of prior works on physical model design~\cite{leishman2014quadrotors, wang2016trajectory, quinonez2020savior, tomic2020simultaneous}:
1) The airframe is rigid and symmetrical. 2) All motors and propellers are rigid and produce the same thrust and torque. Each motor has the same distance to the airframe's \emph{center of gravity} (CoG), leading to a coincidence of CoG with the center of thrust. 3) The Coriolis force and the aerodynamic torque are negligible due to the quadcopter's low flight speed and limited attitude changes.

Our nonlinear physical model is instantiated by deciding the following physical parameters:
\begin{align}
    \label{def:model_params}
    \ModelParams :=& \{
         \VehicleMass, \ArmLength, \InertiaMatrix, \ThrustCoeff, \TorqueCoeff, \TorqueRotorCoeff, \TimeConstant{}, \ThrustSpZero, \ThrustSpRange, \VoltageRef, \ResistanceInternal, \notag \\
         & \MCoef, \BCoefVec \}
\end{align}
\noindent Table~\ref{tab:physical-parameters-definition} presents their definitions in detail. Specifically, $\ThrustCoeff$, $\TorqueCoeff$, $\TorqueRotorCoeff$, $\TimeConstant{}$, $\ThrustSpZero$, $\ThrustSpRange$, $\VoltageRef$, and $\ResistanceInternal$ are related to the propulsion system.
$\MCoef$ and $\BCoefVec$ are related to the aerodynamic drag.
The parameters for each drone type are determined by the measuring and learning procedure in Appendix~\ref{app:Model_Learning}.

The nonlinear physical model $\mathcal{F}$ specifies the relations between the physical parameters $\ModelParams$, the system state $\ModelState(t)$, and the control input $\ModelInput(t)$. With the relation
\begin{align}
 \ModelState(t+1) &= \mathcal{F}_\ModelParams (\ModelState(t),\ModelInput(t))
 \label{def:fitting}
\end{align}
we can predict the runtime system states $\ModelState(t)$.
A precise physical model can accurately predict the drone's motion, improving the effectiveness of attack detection and recovery.
Compared with \savior{}'s physical model \cite{quinonez2020savior}, our physical model specifies the thrust estimation and the aerodynamic drag on the airframe more accurately.
To illustrate our physical model, we concretize $\ModelState(t)$ and $\ModelInput(t)$ into the following states and control inputs:
\begin{align}
    \label{def:model_state}
    \ModelState &= [\pned, \vned, \Aned, \dcm, \AngRateVec, \VoltageLoad, \CurrentLoad, \Windspeed, \rho]^T \\
    \label{def:model_input}
    \ModelInput &= [
        \ThrustSp_1,
        \ThrustSp_2,
        \ThrustSp_3,
        \ThrustSp_4
        ]^T
\end{align}
\noindent All the states in $\ModelState$ and inputs in $\ModelInput$ are timestamped. The system position $\pned$ is the Cartesian coordinates of the CoG. $\vned$ is the linear velocity. $\Aned$ is the linear acceleration.
The system attitude $\dcm$ is a 3-axis rotation matrix, and $\AngRateVec = [\rollrate, \pitchrate, \yawrate]^T$ is the angular velocity specifying how fast the attitude changes along the roll($\roll$), pitch($\pitch$), and yaw($\yaw$) axes.

The relationships between $\pned$, $\vned$, $\Aned$, $\dcm$, and $\AngRateVec$ are well-defined in the \emph{rigid body model}~\cite{tomic2020simultaneous}.
To bridge the relationship between $\ModelState(t)$ and $\ModelInput(t)$, we summarize the rigid body equations as follows:
\begin{align}
    \Aned     &= \Gee \unitvec + \dcm (\CtrlAccel + \DragAccel)  \label{eq:acc_change} \\
     \AngAccVec &= \InertiaMatrix^{-1} (\skewsym{\InertiaMatrix \AngRateVec} \AngRateVec + \CtrlTorque) \label{eq:angacc_change}
\end{align}
\noindent In \eqref{eq:acc_change}, the linear acceleration $\Aned$ is composed of the gravitational acceleration $\Gee$, the acceleration from the control wrench ($\CtrlAccel$), and the aerodynamic drag ($\DragAccel$).
The unit vector $\unitvec = [0, 0, 1]^T$ at the z-axis combines $\Gee$ into $\Aned$, which is located in the North-East-Down (NED) inertial frame.
However, both $\CtrlAccel$ and $\DragAccel$ are located in a different reference frame, the Forward-Right-Down (FRD) body frame.
Therefore, they are multiplied by $\dcm$ to align with the NED frame.
In \eqref{eq:angacc_change}, the angular acceleration $\AngAccVec = [\rollacc, \pitchacc, \yawacc]^T$ is decided by the angular momentum from the airframe ($\skewsym{\InertiaMatrix \AngRateVec_t} \AngRateVec_t$) and the control torque from the actuators ($\CtrlTorque$). $\skewsym{\cdot}$ is the skew-symmetric matrix operator.
$\CtrlAccel$, $\DragAccel$, and $\CtrlTorque$ capture the non-gravitational wrenches applied on the airframe. $\CtrlAccel$ and $\CtrlTorque$ are not measurable during the flight, but they are correlated with the measurable rotor speed $\MotorRotRate_i$ and rotor acceleration $\dot{\MotorRotRate_i},i\in [1...4]$, as shown in \cite{tomic2020simultaneous}.
Specifically, we estimate $\CtrlAccel$ and $\CtrlTorque$ with the following equations:
\begin{align}
    \label{eq:ctrl_accel}
    \CtrlAccel &= \frac{\ThrustCoeff}{\VehicleMass} \left(\MotorRotRate_1^2 + \MotorRotRate_2^2 + \MotorRotRate_3^2 + \MotorRotRate_4^2 \right) \VecStyle{e_3} \\
    \label{eq:ctrl_torque}
    \CtrlTorque &=\left[\begin{array}{l}
        \ThrustCoeff \ArmLength \left(\MotorRotRate_2^2+\MotorRotRate_3^2-\MotorRotRate_1^2-\MotorRotRate_4^2\right), \\
        \ThrustCoeff \ArmLength \left(\MotorRotRate_1^2+\MotorRotRate_3^2-\MotorRotRate_2^2-\MotorRotRate_4^2\right), \\
        \TorqueCoeff \left(\MotorRotRate_1^2+\MotorRotRate_2^2-\MotorRotRate_3^2-\MotorRotRate_4^2\right) + \\
        \qquad \TorqueRotorCoeff \left(\dot{\MotorRotRate_1} + \dot{\MotorRotRate_2} - \dot{\MotorRotRate_3} - \dot{\MotorRotRate_4} \right)
    \end{array}\right]
\end{align}
\noindent Measuring $\MotorRotRate_i$ requires extra sensors.
Therefore, we use the estimated relative thrust $\ThrustEstimate{}_i \geq 0$ as the approximation of $\MotorRotRate_i$.
Given the actuator setpoint $\ThrustSp_i$ of $\ModelInput(t)$, we adjust $\bar{T}_i$ by the battery voltage~\cite{bangura2012nonlinear} and update $\ThrustEstimate{}_i$~\cite{cite_motor_model}:
\begin{align}
    \label{eq:thrust_adjust}
    \ThrustSp^{'}_i(t) &= \frac{\ThrustSp_i(t) - \ThrustSpZero}{\ThrustSpRange} \cdot \frac{\VoltageLoad(t) + \ResistanceInternal \cdot \CurrentLoad(t)}{\VoltageRef} \\
    \label{eq:thrust_delay}
    \ThrustEstimate_i(t) &= \alpha \ThrustEstimate_i(t-1) + (1 - \alpha) \ThrustSp^{'}_i(t)
\end{align}
where $\ThrustEstimate_i(0) = \ThrustSp^{'}_i(0)$. $\alpha = \exp(-\frac{\deltatime}{\TimeConstant})$. $\deltatime$ is the system control interval. $\VoltageLoad$ and $\CurrentLoad$ are the voltage and current measured by the battery power module. Equations~\eqref{eq:thrust_adjust} and \eqref{eq:thrust_delay} provide a more accurate estimation on actual thrust by compensating the voltage drop, mechanical frictions, and the time delay in rotor speed changes.

We estimate the drag-induced acceleration $\DragAccel$ based on~\cite{tomic2020simultaneous}:
\begin{align}
    \label{eq:drag_accel}
    \DragAccel &= \MCoef\Airspeed + \frac{1}{2} \rho \BCoefVec \lVert \Airspeed \rVert \Airspeed
\end{align}
where $\Airspeed = \dcm^T (\vned - \Windspeed)$ is the relative airspeed. $\Windspeed$ is the estimated wind velocity in the NED frame and is obtainable through multi-sensor fusion.
$\rho$ is the air density provided by the barometer. $\lVert \cdot \rVert$ is the vector norm.

\subsection{Online Anomaly Detection}
\label{subsec:method_detection}

Let $\Estimate{\ModelState}(t)$ and $\Measurement{\ModelState}(t)$ be the reference states and sensor measurements, respectively.
Given a sensor instance $i$ of type $j$ that measures $\Measurement{x}_i^j(t)$ on a subset of system states $x^j(t) \subseteq \ModelState(t)$, the residual $r_i^j(t)$ is defined as
    \begin{align}
        \label{def:residual}
        r_i^j(t) &= \Measurement{x}_i^j(t) - \Estimate{x}^j(t)
    \end{align}
\noindent If measurements from this instance significantly deviate from the reference state, the detector raises the alarm to block such a sensor instance from participating in sensor fusion.
State-of-the-art detection approaches \cite{choi2018detecting, choi2020software, quinonez2020savior} take residuals as input and compute a detection statistic $S_i(t)$ to quantify the deviation.
For example, the statistic is defined as the average of the squared residuals~\cite{choi2018detecting} or the cumulative absolute residuals~\cite{choi2020software} within a fixed time window. CUSUM~\cite{woodall1993statistical} used by~\cite{quinonez2020savior} tracks the historical change in residuals and performs better than the time-window detectors~\cite{choi2018detecting, choi2020software} in detecting persistent threats~\cite{urbina2016limiting}.
For each residual $r_i(t)$, CUSUM calculates the detection statistic $S_i(t)$ iteratively based on the following equation:
\begin{align}
    \label{def:stat_cusum}
    S_i(t) &= \max(0, S_i(t-1) + |r_i(t)| - b_i)
\end{align}
where $S_i(0)=0$, and $b_i>0$ is the mean shift to suppress the increment of $S_i(t)$ when there is no attack.
The detector raises the alarm upon $S_i(t)$ exceeding the predefined threshold.

\begin{figure}[t]
    \centering
    \includegraphics[width=0.95\columnwidth]{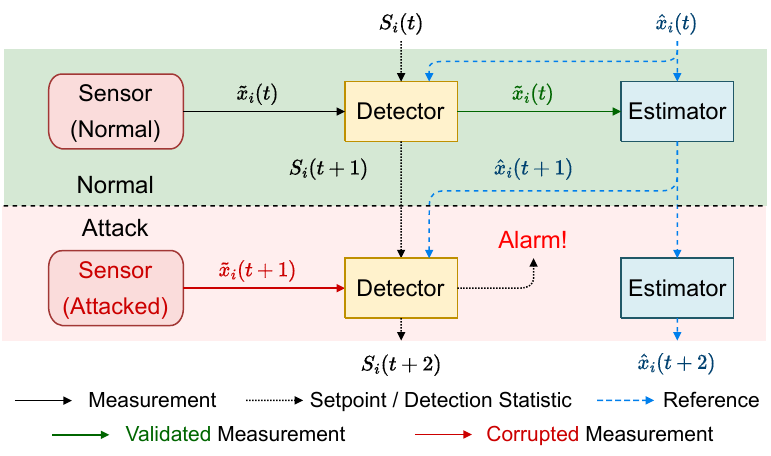}
\vspace{-2ex}
    \caption{Workflow of Anomaly Detection}
    \label{fig:workflow_detection}
\vspace{-2ex}
\end{figure}

However, the specificity of CUSUM would deteriorate when the monitored sensor exhibits intense measurement noise (e.g., white noise, random walk).
The measurement noise affects the sensor data, and a significant presence of noise expands the discrepancy between sensor measurements and reference states (Equation (\ref{def:residual})), leading to an increment of the absolute residual $|r_i(t)|$ used in the CUSUM algorithm.
As a result, CUSUM requires a higher mean shift to suppress the false alarm, allowing the adversary to bypass the detection.
For example, the gyroscope measurements are prone to frequent fluctuations caused by airframe vibrations.
To avoid false alarms, the CUSUM detector on gyroscopes has to be configured with a relatively loose mean shift $b_i$, allowing us to inject a deviation of only 0.04 rad/s to crash the drone while maintaining $|r_i(t)| < b_i$.
According to \eqref{def:stat_cusum}, this deviation keeps the $S_i(t)$ below the detection threshold, and the CUSUM detector will raise no alarm.

To achieve a more robust detection, we propose~\emph{Cumulative Sum-Exponential Moving Average} (CS-EMA) as our detection algorithm.
CS-EMA extends CUSUM with a residual-capped EMA detector, which filters white noise from the input to exhibit the persistent component of deviations:
    \begin{align}
        \label{def:stat_ema_modified}
        S_i(t) &= \lvert MA_i(t) \rvert \\
        \label{def:stat_ema_moving_average}
        MA_i(t) &= \lambda \cdot r^{\prime}_i(t) + (1 - \lambda) \cdot MA_i(t-1) \\
        \label{def:stat_ema_bounded_residual}
        r^{\prime}_i(t) &= \max(\min(r_i(t), +R), -R)
    \end{align}
where $0 < \lambda \leq 1$ scales the impact of newest deviation on the moving average $MA_i(t)$.
Smaller $\lambda$ provides stronger noise filtering ability but requires a longer time to detect the attack.
The \emph{cap} $R$ is a positive value greater than the detection threshold, which defines a hard limit on how much each deviation can impact on the moving average.
Smaller $R$ boosts the relative importance of minor deviations, making the detector focuses on deviations falling into the $[-R,+R]$ interval.
CS-EMA raises the alarm if the $S_i(t)$ produced by the CUSUM component (Equation~(\ref{def:stat_cusum})) or the EMA component (Equation~(\ref{def:stat_ema_modified})) exceeds the alarm threshold.

Our detection approach (\ad{}) applies the CS-EMA algorithm on each sensor instance. An instance may measure more than one system state. For example, GPS measures position and velocity ($\Measurement{x}^\text{GPS} =\{\Measurement{\pned}, \Measurement{\vned}\}$). In that case, each measured state will employ an independent detector. Any alarm from these detectors indicates a compromised sensor instance, and the recovery monitor (\recovery{}) will discharge this instance from the sensor fusion.
Fig.~\ref{fig:workflow_detection} depicts the relationships between the detector, state estimator, and the monitored sensor instance.
This detection workflow is neutral to sensor types.
Each detector responds to every update of its corresponding sensor measurements. Thus, the update rate of $S_i(t)$ follows the sampling rate of the monitored sensor instance.

\subsection{Sensor Isolation and State Recovery}
\label{subsec:method_recovery}

The recovery monitor (\recovery{}) handles the compromised sensor instance identified by the anomaly detectors (\ad{}).
The monitoring procedure comprises two steps in each iteration:~\emph{sensor isolation} (SI) and \emph{system state recovery} (SSR).
When any anomaly detector updates its detection result,
the SI step formulates a validated sensor list by filtering out the compromised sensor instances.
Then, based on a predefined policy, the SSR step picks the best sources from the validated sensor measurements and the reference state. After that, the recovery monitor sends the selected data to the control module and state estimators.

For each system state, the validated sensor measurements determined in the SI step combine with the reference state from \se{} to form an \emph{available data source list}.
The reference state is always treated as an available data source to prevent the interruption of state estimation and flight control.
For example, on a drone with two redundant IMUs, the default available source list of angular velocity $\angrate$ is $\{\Measurement{\angrate}_1^\text{IMU}, \Measurement{\angrate}_2^\text{IMU}, \Estimate{\angrate}^\text{VIMU} \}$ with $\Estimate{\angrate}^\text{VIMU}$ as the reference state from \se{}.
When \ad{} detects the instance $\text{IMU}_1$ is under attack, \recovery{} will remove the source $\Measurement{\angrate}_1^\text{IMU}$~in the SI step, and the available altitude source list becomes $\{\Measurement{\angrate}_2^\text{IMU}, \Estimate{\angrate}^\text{VIMU} \}$.

\begin{figure}[!t]
    \centering
    \includegraphics[width=0.87\columnwidth]{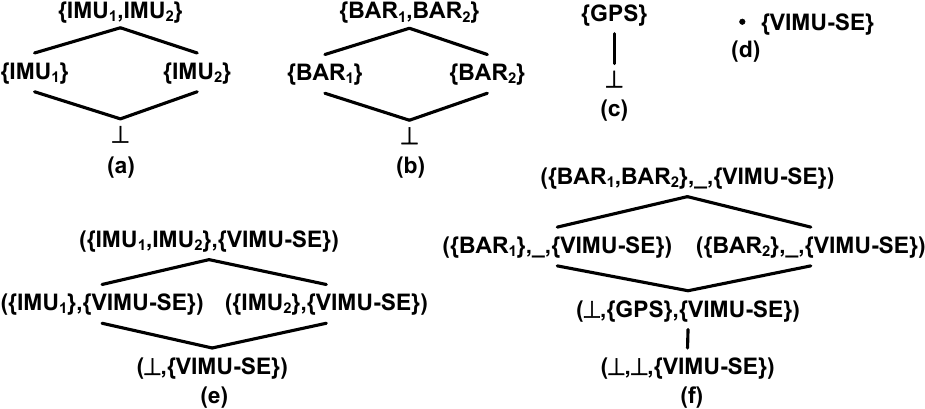}
\vspace{-1ex}
    \caption{Lattice-based Recovery Policy for Flight Altitude ($p_z$) and Angular Velocity ($\angrate$). ``\_'' Stands for \emph{Any Possible Cases} and ``$\bot$'' Stands for \emph{Sensor Instances are Compromised}}
    \label{fig:recovery_lattice}
\vspace{-2ex}
\end{figure}

The SSR step determines which data source in the available data source list is the most reliable and accurate.
Since the quality of data sources varies by the sensor type and the physical model,
we propose a \emph{lattice-based recovery policy} for recovering each system state.
For each system state $x(t)$, we formulate a partial order $(R_{x(t)},<)$ to define the sensor type priority.
For example, $(R_{p_z},<)$ for the flight altitude $p_z$ is \se{}$<$GPS$<$BAR, and $(R_{\angrate},<)$ for the angular velocity is \se{}$<$IMU.
Thus, $(\Measurement{p}_z)^\text{BAR}$ and $\Measurement{\angrate}^\text{IMU}$ have the highest priority in respective case.
We define the compromise lattice of each sensor type, e.g., Fig.~\ref{fig:recovery_lattice}(a)-(d).
Following the priority order relation, we make the ordered multiplications over the compromise lattices to build our lattice-based recovery policy.

The recovery policy lattices for $\angrate$ and $p_z$ are abstracted in Fig.~\ref{fig:recovery_lattice}(e) and (f), respectively.
Each element in Fig.~\ref{fig:recovery_lattice}(e) and (f) represents a \emph{data source list status} under potential sensor attacks.
In each status, \recovery{} chooses the sensor instance from the leftmost valid sensor type (i.e., in best quality).
The recovery monitor is initialized at the uppermost data source list status and will do its best to stay within an upper status in the lattice as long as possible unless the attack identified by \ad{} forces \recovery{} change to a less qualified status in the lattice.
Finally, \recovery{} could reach a status with only \se{}'s reference state as the valid data source.
For example,
when all IMUs are compromised, \recovery{} follows the recovery policy of $\angrate$ to reach the status ($\bot$,\{\se{}\}) of Fig.~\ref{fig:recovery_lattice}(e), which means the control module receives the uninterrupted reference angular velocity $(\Estimate{\angrate})^\text{VIMU}$ for flight control.

\subsection{State Estimator and Buffer Safeguard}
\label{subsec:method_ekf}

\begin{figure}[!tb]
    \centering
    \includegraphics[width=0.9\linewidth]{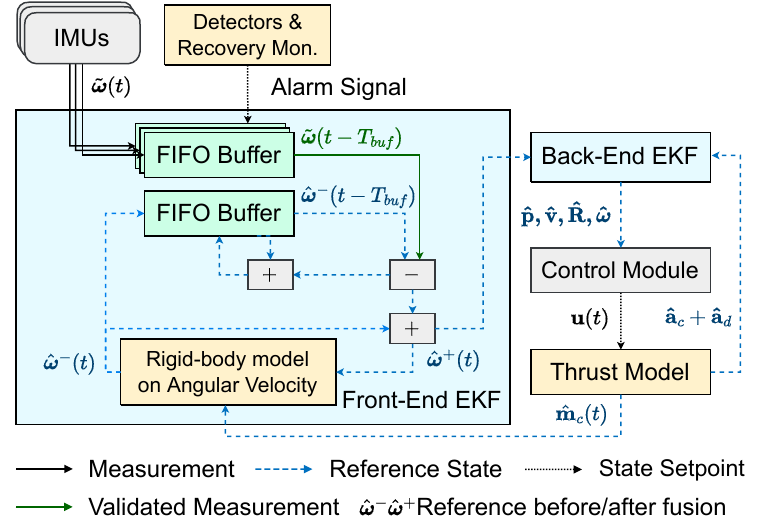}
\vspace{-1ex}
    \caption{Workflow of Front-End EKF of \se{}}
    \label{fig:workflow_estimator}
\vspace{-2ex}
\end{figure}

Our fine-grained physical model can accurately capture the transition of the system states.
However, due to the model bias and external perturbation, the model prediction could still slightly deviate from the actual system state.
Such deviation can accumulate over time during the flight.
Therefore, \sysname{}'s state estimator (\se{}) uses a pair of \emph{Extended Kalman Filters} (EKF) in a cascaded pattern to periodically rectify the deviation with benign measurements from the operational sensors.
The \emph{front-end EKF} fuses the current angular velocity $\Estimate{\angrate}^-$ (predicted by \eqref{eq:angacc_change}) with IMU measurements $\Measurement{\angrate}$ to produce the corrected estimate $\Estimate{\angrate}^+$. Then, the \emph{back-end EKF} updates the reference states with $\Estimate{\angrate}^+$ and available measurements of position, speed, and attitude.
This process further rectifies $\Estimate{\angrate}^+$ with other sensor data to derive the reference angular velocity $\Estimate{\angrate}$ given to the anomaly detectors and recovery monitor.

The back-end EKF uses a structure similar to the vanilla state estimator in the autopilot.
However, to address IMU attacks, the design of the front-end EKF is nontrivial (Fig.~\ref{fig:workflow_estimator}).
An advanced adversary can delay the alarm to several hundred milliseconds after the attack. Considering the high sampling rate of IMU sensors (from a hundred to thousands hertz), such a time delay allows the attacker to inject a remarkable bias into the reference state, rendering detection and recovery inoperable~\cite{jang2023paralyzing, gao2023exploring}.
To address this threat, we develop first-in-first-out (FIFO) buffers in the front-end EKF as the safeguard between the recent measurements and angular velocity estimates: 1) \emph{Measurement buffers}: The front-end EKF holds one \emph{measurement buffer} for each IMU instance to introduce a time delay to the sensor measurements. 2) \emph{Estimate buffer}: The \emph{estimate buffer} raises a similar time delay to align the timing of estimates with the delayed sensor measurements for a correct sensor fusion.

Even if the alarm has been delayed, such buffers can prevent the faulty data from being immediately fused into the estimation.
At each iteration of state estimation, the front-end EKF first checks the detection report from \recovery{}.
If \recovery{} reports a compromised IMU instance, the EKF will remove all IMU measurements stored in the corresponding buffer to prevent the potential attack.
Then, the EKF updates the estimation with \eqref{eq:angacc_change}, pushes the newest estimate into, and pops the oldest estimate from the estimate buffer. After that, the EKF iterates over the measurement buffers to find a validated and time-aligned IMU measurement.
Such a measurement can be unavailable due to the sensor attack. In that case, the front-end EKF will output $\Estimate{\angrate}^-$ to the back-end EKF and let it correct the estimate with other sensors. Otherwise, it pops the IMU measurement out from the buffer to perform sensor fusion.
The sensor fusion in each iteration proceeds as follows: the EKF calculates a residual between the delayed estimate and measurement, adds the residual to the newest estimate $\Estimate{\angrate}^-(t)$, and outputs the corrected estimate $\Estimate{\angrate}^+(t)$.
The same residuals are also added to the estimates in the estimate buffer to cumulate the corrections over time.
The buffer size is determined by:
\begin{align}
    \label{def:buf_size}
    size_\text{buffer} &:= 1 + \lceil \TBuffer \cdot H \rceil
\end{align}
where $\TBuffer$ is the desired time to hold the received measurements, and $H$ is the output rate of the stored data.

With the FIFO buffer safeguard, we can further strengthen the integrity of reference states even under stealthy IMU attacks.
However, using our buffer safeguard could increase the bias in reference angular velocity because of the delayed sensor fusion. Such bias causes a modest reduction in the detector specificity, especially for CUSUM.
Our CS-EMA algorithm does not rely on the CUSUM component to identify stealthy attacks.
Therefore, we add an extra reference angular velocity for the CS-EMA detectors that monitor the gyroscope. This reference state is received by the CUSUM component of CS-EMA and is corrected by the newest IMU measurements to shorten the detection delay towards overt attacks. As both improvements work independently, the extra reference improves the overall performance of the CS-EMA detector while maintaining its robustness towards stealthy attacks.

\section{Implementation}\label{sec:implementation}

\subsection{Security-Enhanced Autopilot}

We implement \sysname{} on PX4 v1.13.3.
We add new message formats to the PX4 architecture for reference state and sensor health status.
Then, we implement the detection algorithms as a library, including CS-EMA, CUSUM, and the time-window detections~\cite{choi2018detecting, choi2020software}.
We insert \ad{} into the sensor middlewares and modules to ensure the detection covers all received measurements.
The detector location depends on the sensor type.
For example, we insert the detectors for the gyroscope and accelerometer into the middleware \texttt{PX4Gyroscope} and \texttt{PX4Accelerometer}.
For other sensors, we implement detectors in their data processing modules (e.g., \texttt{vehicle\_gps\_position}).
We implement \recovery{} in the state estimation (\texttt{EKF2}) and estimator selector (\texttt{EKF2Selector}) modules.
\revise{The recovery monitor receives the runtime detection results. Its IMU-related component in \texttt{EKF2Selector} applies the IMU-related recovery policies (e.g., Fig.~\ref{fig:recovery_lattice}(e)) to isolate compromised IMUs by orderly responding to the angular velocity, acceleration, and attitude errors. Similarly, \recovery{}'s IMU-unrelated component in \texttt{EKF2} applies the IMU-unrelated policies (e.g., Fig.~\ref{fig:recovery_lattice}(f)) to isolate other sensors by orderly responding to the horizontal position, altitude, and velocity errors.}
We incorporate the physical model into the two cascaded EKFs of \se{}. For angular velocity estimation, we add a new module (\texttt{VirtualIMU}) to the autopilot. This module integrates the front-end EKF and the nonlinear physical model. It uses actuator setpoints from the \texttt{actuator\_outputs} message to update current acceleration and angular velocity. For the back-end EKF, we reuse the existing EKF2 module as it already provides estimation and sensor fusion in the position, velocity, and attitude states. When all hardware IMUs are compromised, \texttt{EKF2Selector} will switch to the \se{} for state estimation, thus preventing the in-flight control loss leading to drone crashes.

\subsection{Sensor Attack Simulation}
\label{sec: attack implementation}
\label{subsec:attack_implementation}

\begin{table}[!tb]
\renewcommand{\arraystretch}{1.35}
\caption{Sensor Attacks Used in Our Evaluation}
\vspace{-1ex}
\footnotesize
\label{table:attack_types}
\def \DescWidthTableAttackType {6.2cm}
\resizebox{\columnwidth}{!}{
            \begin{tabular}{p{1cm}<{\centering} p{1cm}<{\centering} p{\DescWidthTableAttackType}}
            \hline
            \textbf{Category}               & \textbf{Notation} & \textbf{Description} \\
            \hline
            \multirow{10}{*}{Overt}      & $\OAGps$          & Injecting spoofed GPS positions \cite{choi2018detecting, choi2020software, quinonez2020savior}           \\ \cline{2-3}
                                        & \multirow{2}{*}{$\OAGpsJoint$}     & Combines position and velocity spoofing to deviate a hovering drone with faked GPS signal~\cite{sathaye2022experimental} \\
                                        \cline{2-3}
                                        & $\OATwoGyro$      & \multirow{3}{\DescWidthTableAttackType}{Modulated acoustic attacks injecting controlled fixed deviation to target gyroscopes or accelerometers~\cite{gao2023exploring}.\\ The gyroscope attacks follow the scenarios in \cite{choi2020software}.} \\
                                        & $\OAThreeGyro$  \\
                                        & $\OAThreeAccel$  \\
                                        \cline{2-3}
                                        & \begin{tabular}[t]{@{}c@{}} $\OAGyroModel{3}{ICM20602}$ \\ $\OAGyroModel{3}{ICM20689}$\end{tabular} & Unmodulated acoustic attacks injecting sinusoidal attack signal into gyroscope measurements of two IMU models: ICM20602 and ICM20689~\cite{tu2018injected, jeong2023unrock} \\
                                        \cline{2-3}
                                        & $\OABaro$ & Injecting fixed deviations to all barometers \cite{choi2020software} \\
                                        \cline{2-3}
                                        & $\OAMag$ & Injecting fixed deviations to all magnetometers \cite{nashimoto2018sensor} \\
            \hline
            \multirow{2}{*}{Stealthy}   & $\SAGps$          & Stealthy position spoofing              \cite{quinonez2020savior}  \\
            \cline{2-3}
                                        & $\SAGyro$         & Stealthy gyroscope attack \cite{gao2023exploring} following the scenario of \cite{quinonez2020savior}  \\
            \hline
            \multirow{3}{*}{Multi-Type} & \multirow{3}{*}{$\MSAttack{*}$}    & A combination of multiple overt attacks~\cite{nashimoto2018sensor, choi2020software, attack-hardness-oakland24} for testing recovery effectiveness. Especially, attack $\text{MA}_{\text{Mag}|\text{Accel}|\text{Gyro}}$ is equivalent in impact to the EMI attack of~\cite{jang2023paralyzing} according to the implementation of~\cite{attack-hardness-oakland24} \\
            \hline
            \end{tabular}
}
\vspace{-3ex}
\end{table}

Following~\cite{choi2018detecting, choi2020software, quinonez2020savior}, we simulate sensor attacks by instrumenting attack code into the sensor middleware and modules of the autopilot.
These middlewares and modules convert raw sensor readings (e.g., pressure, currents, and magnetic field strength) into measurements of system states
(e.g., position, acceleration, and attitude) before delivering them to the state estimators and control module.
\revise{
Modifying those measurements within the autopilot to imitate the impact of injected attack signals has also been adopted in~\cite{fei2020learn, zhang2020real, zhang2021real}. These software-based attacks also allow us to efficiently apply attacks on various sensors with diverse attack amplitudes.
}

The number and kinds of victim sensors vary by the attack type.
Table~\ref{table:attack_types} presents the sensor attacks used in our evaluations, including all sensor attacks from baseline approaches~\cite{choi2018detecting, choi2020software, quinonez2020savior} and several recent popular attacks \cite{sathaye2022experimental, gao2023exploring, tu2018injected, nashimoto2018sensor, jang2023paralyzing}.
We denote the attack in the form of $\textit{attack\_category}_\textit{sensor\_type}^{\text{\#}\textit{compr}/\text{\#}\textit{avail}}(\textit{deviation})$.
Specifically, we implement overt attacks (OA), stealthy attacks (SA), and multi-type attacks (MA) on different sensor types (GPS, Gyro, Accel, Baro, or in combination). \#\emph{compr} and \#\emph{avail} respectively represent the number of compromised and available sensor instances, which are omitted if only one instance exists in the drone. The \emph{deviation} specifies the amplitude of injected deviation to the compromised instances, which could be constant or time-dependent. More specifically, the adversary of the modulated acoustic attacks spoofs the majority ($\OATwoGyro$) or all ($\OAThreeGyro$) of the gyroscopes, misleading the drone's majority voting to discharge the operational gyroscope or even disabling all redundant IMUs.
The unmodulated acoustic attacks ($\OAGyroModel{3}{ICM20602}, \OAGyroModel{3}{ICM20689}$) follow the principle of \cite{tu2018injected} to compromise our quadcopter's gyroscopes ICM20602 and ICM20689. The time-dependent deviation is defined as $A_{i} \cdot \cos{(2 \pi F_{i} \cdot t)}$,
where $t$ is the time elapsed since the attack, $A_i \in (0, A_{max}]$ is the induced amplitude ($A_{max}$ is the maximum induced amplitude), and $F_{i}$ is the induced acoustic frequency. $A_{max}$ and $F_{i}$ are decided by the targeted sensor model. According to~\cite{jeong2023unrock}, we have $A_{max}$= 0.927 rad/s and $F_i$= 19.7 Hz for $\OAGyroModel{3}{ICM20602}$; $A_{max}$= 1.899 rad/s and $F_i$= 205.9 Hz for $\OAGyroModel{3}{ICM20689}$, respectively.

The overt attacks represent the common attack scenarios.
However, the most threatening attacks against PBAD are the stealthy attacks~\cite{quinonez2020savior, urbina2016limiting}.
Stealthy attackers know the victim drone's characteristics, e.g., mission plan, reference states, detection algorithm and parameters.
Since \sysname{} and the related PBAR approach~\cite{choi2020software} require a prior detection alarm to initialize the recovery procedure, stealthy attacks can prevent or delay the recovery effects, causing more significant damage.
The stealthy attacks $\SAGps$ and $\SAGyro$ in Table~\ref{table:attack_types} adaptively inject a minor deviation to all target sensor instances, maximizing the attack effect without triggering the alarm.
Specifically, the attacker tries injecting a series of attack signals $\tilde{x}^{*}_{i}(t)$ into the targeted sensor instance $i$ to maximize the deviation misleading the system behaviors while holding the detection statistic $S_i(t)$ under the alarm threshold $\tau$:
\begin{align}
    \label{eq:attack_residual}
    r^*_i(t) =\Measurement{x}_i(t) + \Measurement{x}^*_i(t)-\Estimate{x}_i(t) \\
    \label{def:stealthy_attack}
    \operatorname*{argmax}_{\Measurement{x}^{*}_{i}(t)} |r^*_i(t)|\  s.t.\ S_i(t) \leq \tau_i
\end{align}
where $r^*_i(t)$ is the residual that includes $\Measurement{x}^*_i(t)$.
The most effective $\Measurement{x}^{*}_{i}(t)$ depends on the detector parameters and the difference between measurement $\Measurement{x}_i(t)$ and reference state $\Estimate{x}_i(t)$.
Finally, the multi-type attacks in Table~\ref{table:attack_types} apply all-instance overt attacks on multiple sensor types simultaneously. Following the setup of~\cite{attack-hardness-oakland24}, we use the attack $\text{MA}_{\text{Mag}|\text{Accel}|\text{Gyro}}$ to simulate the impact of the EMI attack~\cite{jang2023paralyzing} on the simulator.

\section{Evaluation}\label{sec:evaluation}

\subsection{Experimental Setup}\label{subsec:setup}

\subsubsection{Testbed}\label{subsubsec:testbed}

We evaluate \sysname{} with quadcopters in simulation and real-world flight. We use the default simulator of PX4 autopilot, i.e., jMAVSim, and its default quadrotor as the simulation testbed.
Wind is a common external disturbance in flight. Thus, we evaluate the robustness of \sysname{} with the wind simulation enabled (more results in Appendix~\ref{app:evaluation-wind}). We also activate all the detectors throughout experiments rather than only those of the sensors under attack. This setting can reveal how the recovery performance is affected by detection delay and false alarms.
For the real-world evaluations, we assemble a quadcopter with a ZD550 airframe. The autopilot board is a CUAV V5+, with a 216MHz Arm Cortex-M7 CPU and 512 kB RAM for runtime memory. We calibrate all sensors before real-flight tests, and Table~\ref{table:sensor_list} lists the sensors used by this drone. Throughout the evaluation, we use QGroundControl as the ground control station and communicate with the drone through the Mavlink v2.0 protocol.

\begin{table}[!tp]
\renewcommand{\arraystretch}{1.2}
\caption{List of Sensors of Real World Drone}
\label{table:sensor_list}
\centering
\scriptsize
\vspace{-1ex}
    \begin{tabular}{c c c c}
        \hline
        \textbf{Sensor Type} & \textbf{Product Type} & \textbf{Number} & \textbf{Measurement} \\
        \hline
        GPS Module & Neo V2 & 1 & Position and Velocity \\ \hline
         & ICM20602 & 1 & Acceleration \\ \cline{2-3}
        IMU & ICM20689 & 1 & and \\ \cline{2-3}
         & BMI055 & 1 & Angular Velocity \\ \hline
        Magnetometer & IST8310 & 2 & Attitude \\ \hline
        Barometer & MS5611 & 2 & Vertical Position \\
        \hline
    \end{tabular}
\vspace{-3ex}
\end{table}

\subsubsection{Baselines}

We compare \sysname{} with the following approaches, including the variants of \sysname{} and the state-of-the-art PBAD \cite{choi2018detecting, quinonez2020savior} and PBAR \cite{choi2020software} approaches.
\begin{compactitem}[-]
  \item \emph{Control Invariant}\footnote{\url{https://github.com/wangwwno1/Project-VIMU/tree/baseline/CI}} (\ControlInvariant)~\cite{choi2018detecting} uses a linear state-space model to obtain the expected system state and applies a squared error time-window detector (denoted as L2TW).

  \item \emph{Software-based Real-time Recovery}\footnote{\url{https://github.com/wangwwno1/Project-VIMU/tree/baseline/SRR}} (\SoftwareSensor{})~\cite{choi2020software} reuses the linear model for recovery and applies a time-window detector based on the absolute error (denoted as L1TW).

  \item \emph{\savior{}}\footnote{\url{https://github.com/wangwwno1/Project-VIMU/tree/baseline/SAVIOR}\label{fn:savior}}~\cite{quinonez2020savior} predicts system states with a nonlinear physical model and applies CUSUM for detection.

  \item \emph{\saviorbuffer{}}\footref{fn:savior} is a PBAR approach that extends \savior{} with the recovery monitor (Section~\ref{subsec:method_recovery}) and FIFO-buffer guarded state estimation (Section~\ref{subsec:method_ekf}). It also delivers the estimated state as an available data source for the SSR step in Section~\ref{subsec:method_recovery}.
  \item \emph{\vimucusum{}}\footnote{\url{https://github.com/wangwwno1/Project-VIMU/tree/baseline/Virtual-IMU+CUSUM-Detector}} is a variant of \sysname{} that replaces CS-EMA with the CUSUM detector to demonstrate the contribution of our detector.

  \item \emph{\vimuNoBuffer{}}\footnote{\url{https://github.com/wangwwno1/Project-VIMU}} is a variant of \sysname{} that discharges FIFO buffers to demonstrate the contribution of buffer safeguard. This variant's reference state is corrected with the newest IMU measurements.

\end{compactitem}
The original implementations of \cite{choi2018detecting, choi2020software, quinonez2020savior} are on different autopilots and simulators. Therefore, we reimplement these baselines on PX4 autopilot v1.13.3 to ensure a fair comparison.
We obtain CS-EMA detector parameters and follow the procedures described in \cite{choi2018detecting, choi2020software, quinonez2020savior} to obtain the model and detector parameters (Appendix~\ref{app:Param_Selection}).
Comparing \sysname{} with machine-learning-based approaches, i.e., \cite{DBLP:conf/dsn/DashLCKP21, fei2020learn, jeong2023unrock}, is out of our scope due to their high computational overheads and the difficulty of deployment over low-end drones.

\subsubsection{Data Collection}\label{subsec:data_collection_eval}

We design \revise{three} missions to represent the most common flight states of drones. 1) \textit{Hovering}: Taking off to an altitude of 15m before flying horizontally to a preset waypoint located 10m north and 10m east of the launch point. Then, hovering for 300 seconds before returning to the home. 2) \textit{Moving}: Taking off to an altitude of 50m before flying horizontally to the same preset waypoint. Then, flying horizontally to the northeast for a distance of 1,000m before returning to the home.
\revise{ 3) \textit{Maneuver}: Taking off to an altitude of 15m. Then, flying horizontally in an equilateral-triangle route with an edge length of 2.5m. On this route, the drone changes its direction between clockwise and counter-clockwise for ten times whenever reaching the initial hovering point.
In the flight, we randomly pick one waypoint from the flight direction changing point or the triangle vertices, as the preset waypoint.
In Section~\ref{sec:evaluation}, we mainly present \sysname{}'s effectiveness with the \textit{Hovering} and \textit{Moving} scenarios. We discuss the impact of drastic maneuvers (e.g., rapid acceleration and sharp turns in the \textit{Maneuver} scenario) on \sysname{}'s effectiveness in Appendix~\ref{app:evaluation_maneuver}.}
We combine the mission scenarios with the attacks listed in Table~\ref{table:attack_types}, in terms of $\textit{Mission}[Attack]$, to represent the attack test cases.
For example, $\textit{Hovering}[\OAGps(10.0)]$ is an overt GPS attack test case with the attack deviation set to 10.0 meters in the \textit{Hovering} scenario.

We compare \sysname{} with the baselines on various attack test cases. For each candidate approach, we collect 50 flight records per attack test case.
In the test case, the drone initializes the system, takes off, and activates the detectors before reaching the preset waypoint.
After the drone reaches the waypoint, we activate the attack and record the attack-beginning timestamp $t_\text{Atk}$.
The anomaly detectors will examine received measurements to decide whether the monitored sensor instance $s(i)$ is under attack.
If the detector of $s(i)$ alarms (including false alarms), we record the first-alarm timestamp on instance $s(i)$, i.e., $t_\text{Alarm}^{s(i)}$. We define the earliest $t_\text{Alarm}^{s(i)}$ of all sensor instances as the first-alarm timestamp of this flight, i.e., $t_\text{Alarm}$.
Regardless of the detection results, the drone continues the flight until it completes its mission, crashes, or the position deviation between the estimate $\hat{\pned}(t)$ and the actual position $\pned(t)$ has reached 5 meters.
We group the collected records by sensor type after the flight ends.

\subsubsection{Metrics}

We evaluate detection effectiveness with \emph{true positive rate} (TPR), \emph{false positive rate} (FPR), and \emph{time to detect} (TTD).
TPR and FPR are classical metrics in anomaly detection.
They measure the sensitivity and specificity of the detectors.
The attacks targeting IMUs can cause irreparable damage within tens to hundreds of milliseconds~\cite{jeong2023unrock, jang2023paralyzing}, and GPS spoofing requires several seconds to become effective~\cite{sathaye2022experimental, shen2020drift}. Therefore, we define an upper time-bound $T_\text{Alarm}$ on the detection delay, e.g., $T^\text{GPS}_\text{Alarm} = 20$ s and $T^\text{Gyro}_\text{Alarm} = 1$~s.
When deciding the detection metrics, we consider the alarm \emph{effective} in attack mitigation if and only if $t_\text{Atk} \leq t_\text{Alarm}^{s(i)} \leq t_\text{Atk}+T_\text{Alarm}$.
Given a sensor instance participated in a flight mission, TPR and FPR are decided on the following detection result classification:
\begin{compactitem}[-]
    \item \emph{True positive} (TP): The detector raises an effective alarm when the monitored instance is compromised.
    \item \emph{False positive} (FP): The detector raises the alarm despite the instance being operational.
    \item \emph{True Negative} (TN): The detector correctly raises no alarm on an operational instance.
    \item \emph{False Negative} (FN): The detector fails to raise an effective alarm on a compromised instance, i.e., either raising no alarm or raising an ineffective alarm later than the upper time-bound $T_\text{Alarm}$.
\end{compactitem}

\noindent \revise{Note that each detection result and its classification are defined on the flight mission.}
Then, the \emph{time to detect} is the time a detector takes to identify the presence of the attack.
A shorter TTD implies a faster response in attack detection and isolation.
As the recovery procedure starts with attack isolation, the TTD estimates the attack damage before detection and affects the recovery performance.
For a sensor instance $s(i)$ under attack, if the detection result is TP, we calculate TTD with $(t_\text{Alarm}^{s(i)} - t_\text{Atk})$.
Otherwise, we label TTD as $\geq T_\text{Alarm}$ to indicate that the detector failed to report the attack within a reasonable time.

Following \SoftwareSensor{}~\cite{choi2020software}, we use the \emph{effective recovery duration} as the metric of recovery performance.
We regard the recovery procedure as \emph{in effect} if any detector raises the alarm and, since then, the distance $d(t)$ between the current position estimate $\hat{\pned}(t)$ and the actual position $\pned(t)$ never exceeds the error threshold $\epsilon$.
\begin{align}
    \label{def:recovery_in_effect}
    d(t) = ||\pned(t) - \hat{\pned}(t)|| \leq \epsilon, t \in [1, ..., k]
\end{align}
where $t$ is the timestamp of each position state, $k$ is the first timestamp of $d(t) > \epsilon$ or the end of the flight. Thus, we define \emph{effective recovery duration} as the time from the first alarm ($t_\text{Alarm}$) to the above timestamp $k$. We set $\epsilon$ as 3 meters throughout the evaluations.
Due to the flight time constraint, we cannot always measure the maximum recovery duration since in several attack categories, the recovery approaches can sustain the flight for a long time. We set an upper time bound $T_{rec}=300$~s for the effective recovery duration, and label the result as $\geq T_{rec}$ if the approach has an effective recovery duration longer than $T_{rec}$.

\subsection{Detection Effectiveness}

We first evaluate the detection effectiveness of the candidate approaches using overt attacks, including GPS spoofings ($\OAGps$, $\OAGpsJoint$) and gyroscope attacks ($\OATwoGyro$, $\OAThreeGyro$, $\OAGyroModel{3}{ICM20602}$, $\OAGyroModel{3}{ICM20689}$).
Then, we evaluate the resilience of different approaches toward stealthy attacks ($\SAGps$, $\SAGyro$).
The measurements affected by GPS spoofing are the north-axis position (in meters) and velocity (in m/s, only for $\OAGpsJoint$). For gyroscope attacks, the injected signal applies to the three-axis angular velocity (in rad/s).

\subsubsection{Effectiveness on Overt Attacks}
\label{subsec:effectiveness_on_overt_attacks}

\begin{figure}[tb!]
\centering
    \includegraphics[width=0.8\linewidth]{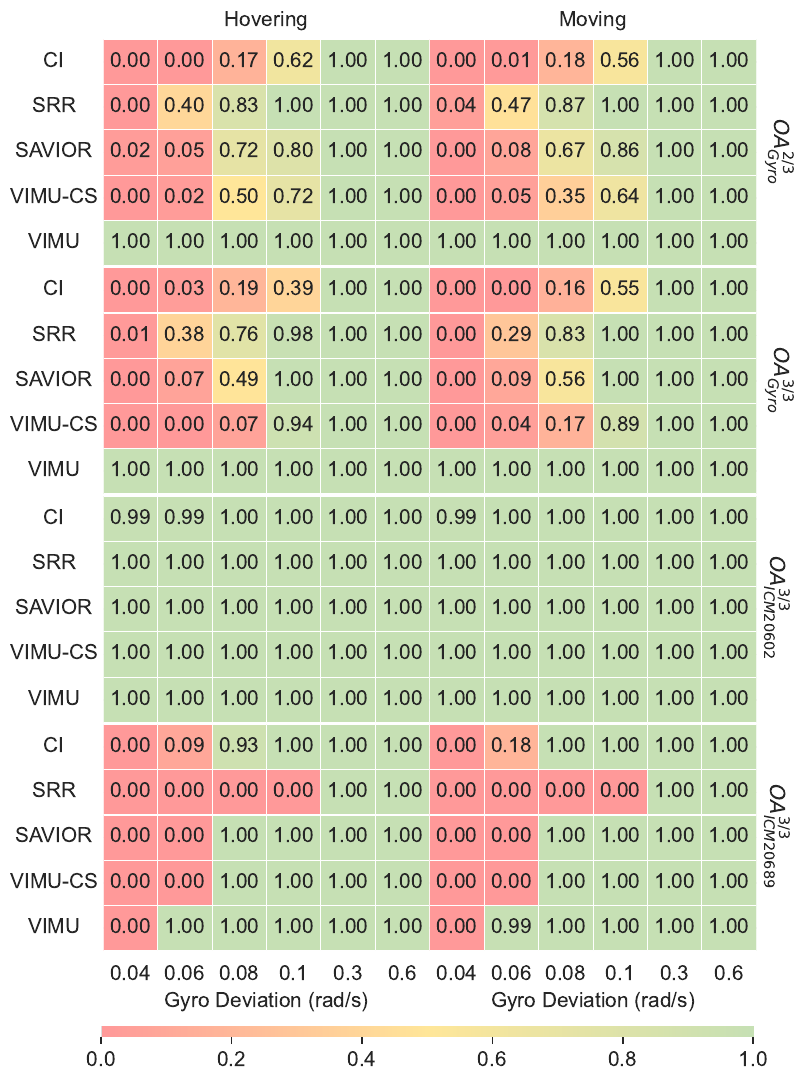}
\vspace{-2ex}
    \caption{TPR Heatmap on Overt Gyro Attacks}
    \label{fig:tpr_heatmap}
\vspace{-2ex}
\end{figure}

\begin{figure}[tb!]
\centering
    \includegraphics[width=0.95\linewidth]{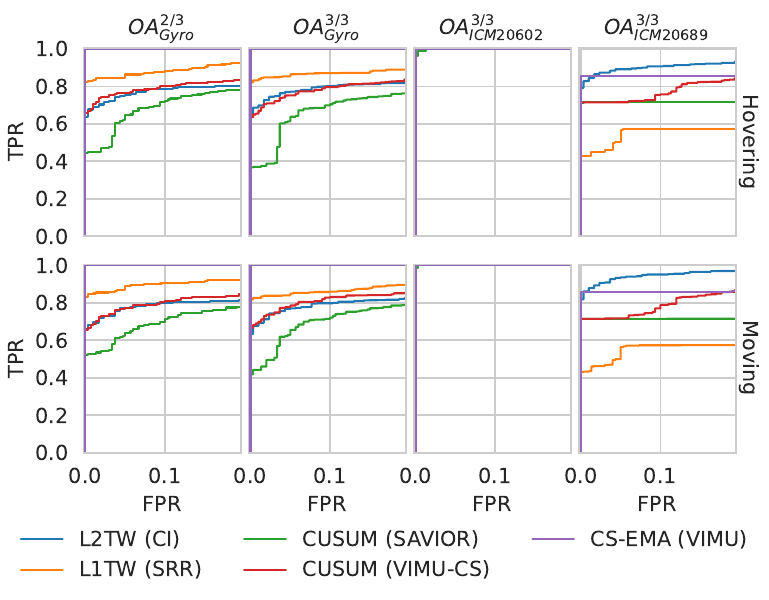}
\vspace{-2ex}
    \caption{ROC Curves on Overt Gyro Attacks}
    \label{fig:roc_curves}
\vspace{-3ex}
\end{figure}

Fig.~\ref{fig:tpr_heatmap} shows that the TPR
positively correlates with the attack deviation of overt gyroscope attacks.
All the approaches achieve 100\% TPR when the attack deviation reaches 0.3 rad/s.
In almost all attack test cases, \sysname{} has the highest TPR and the largest area under the ROC curves (Fig.~\ref{fig:roc_curves}).
Our CS-EMA detector successfully detects most attacks with minor deviations (0.04 rad/s and 0.06 rad/s), whereas the CUSUM detectors of \savior{} and \vimucusum{} achieve lower TPRs.
Such attacks have deviations considerably smaller than the angular velocity encountered in quadcopter's flight maneuvers (up to 2.0 rad/s~\cite{choi2020software}), but can lead to control loss within hundreds of milliseconds.
Therefore, our CS-EMA detector is more suitable for identifying threats targeting the gyroscope.
Compared with $\OAGyroModel{3}{ICM20602}$, detecting $\OAGyroModel{3}{ICM20689}$ is more challenging because its attack signal operates at a frequency ($F_i = 205.9$ Hz) closer to the angular velocity control (250 Hz in our evaluation), which makes it more similar to the measurement noise described in Section~\ref{subsec:method_detection}.
We further compare the AUC of \vimucusum{} and \savior{} in Fig.~\ref{fig:roc_curves} to determine the contribution of the physical model to detection effectiveness.
As the FPR is calculated on attack-free flight records, it will not be affected by the FIFO buffer (Section~\ref{subsec:method_ekf}).
Meanwhile, both solutions use the same CUSUM detector.
Thus, the difference in AUC mainly comes from the fine-grained physical model (Equations~\eqref{eq:ctrl_torque} to ~\eqref{eq:drag_accel}), indicating that the fine-grained physical model has contributed to attack detection.
GPS data is less affected by the airframe vibration than the gyroscope data. Therefore, our CS-EMA detector performs similarly to the CUSUM of \savior{} and \vimucusum{} in detecting $\OAGps$.
The implementation of $\OAGpsJoint$ does not have a constant deviation setting to elaborate on the TPRs on various attack deviations. We observed that \sysname{}, \savior{}, and \vimucusum{} can detect $\OAGpsJoint$ in 100\% TPR with an AUC=0.98$\sim$1.00, while \ControlInvariant{} and \SoftwareSensor{} can only achieve an AUC less than 0.15.

\begin{figure}[tb]
    \centering
    \includegraphics[width=0.9\linewidth]{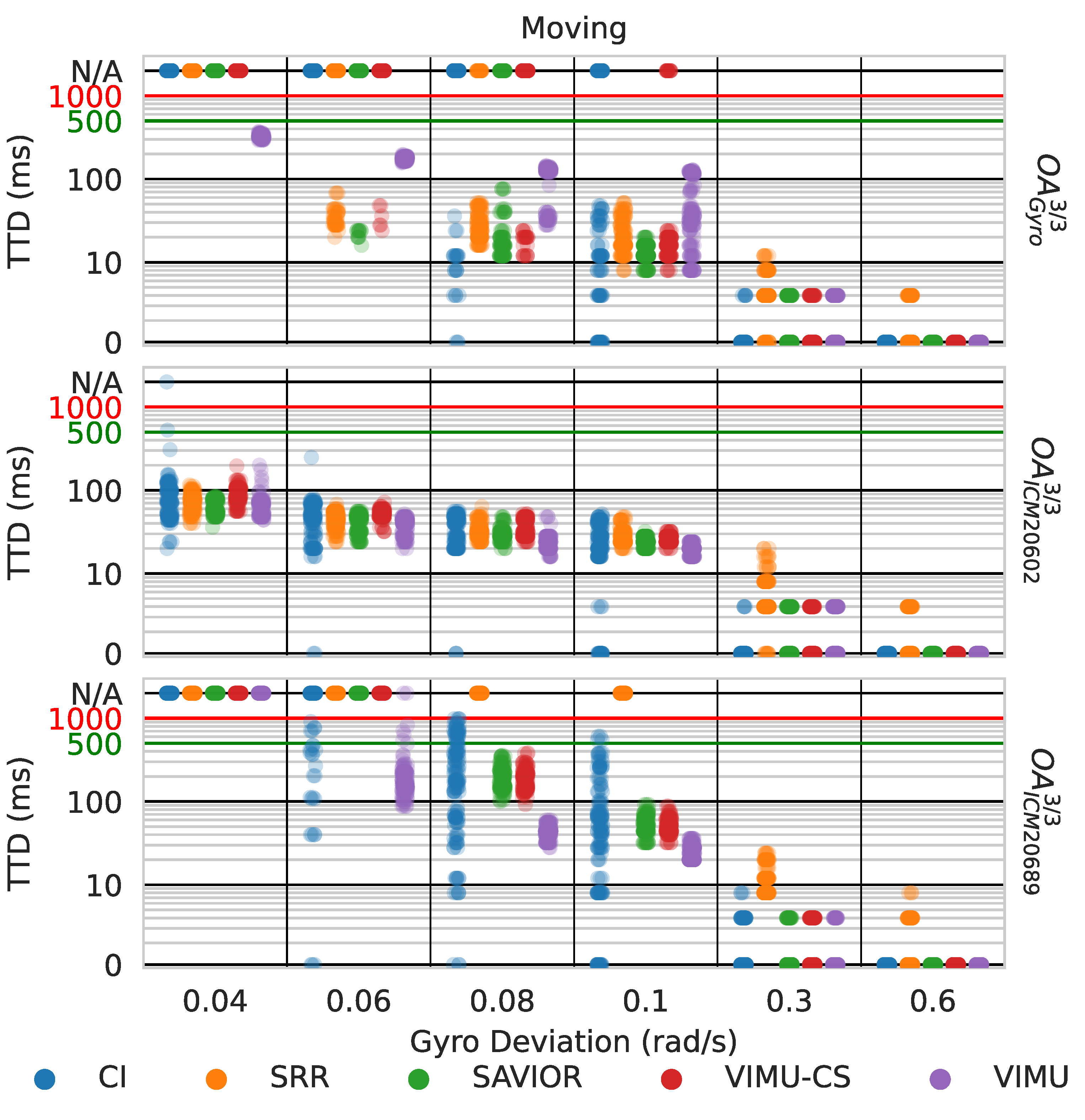}
\vspace{-2ex}
    \caption{Time to Detect on Overt Gyro Attacks in \textit{Moving} Mission (Red Line: $T^\text{Gyro}_\text{Alarm}$; Green Line: $\TBuffer$)}
    \label{fig:time_to_detect_gyro_moving}
    \label{fig:time_to_detect}
\vspace{-3ex}
\end{figure}

\begin{figure}[tb!]
    \centering
    \includegraphics[width=0.9\linewidth]{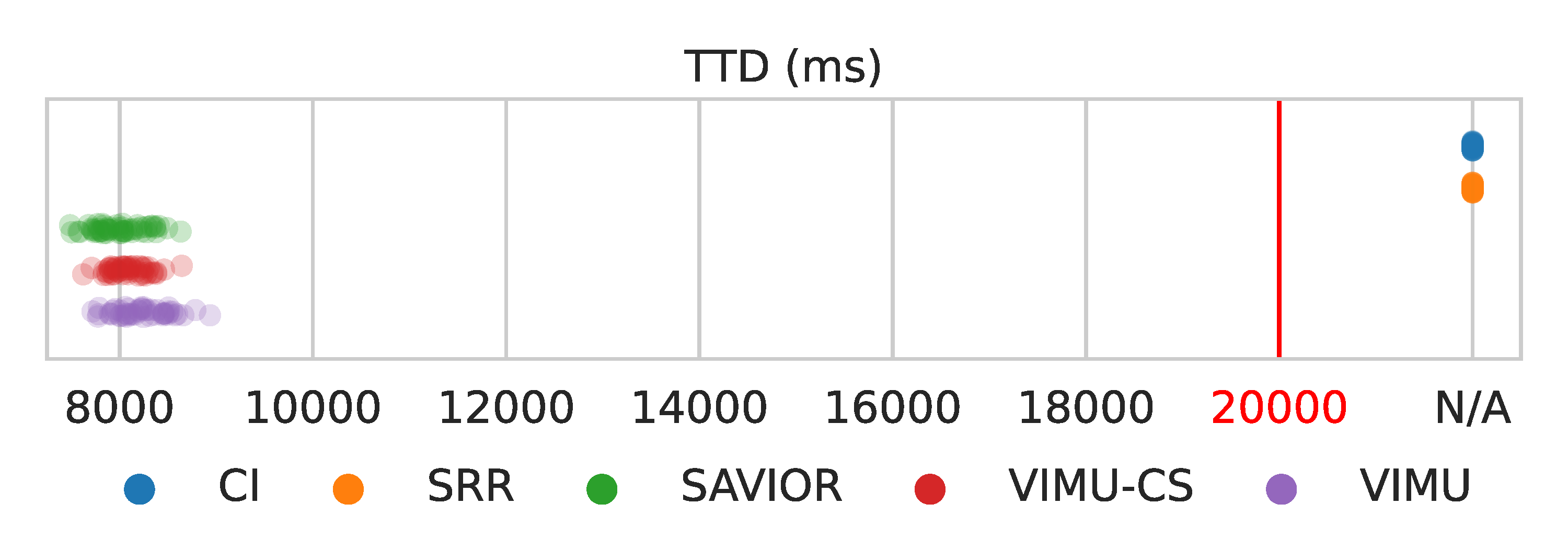}
\vspace{-2ex}
    \caption{TTD on GPS Spoofing $\OAGpsJoint$}
    \label{fig:gps_joint_attack}
\vspace{-3ex}
\end{figure}

\begin{figure}[tb!]
    \centering
    \includegraphics[width=0.9\linewidth]{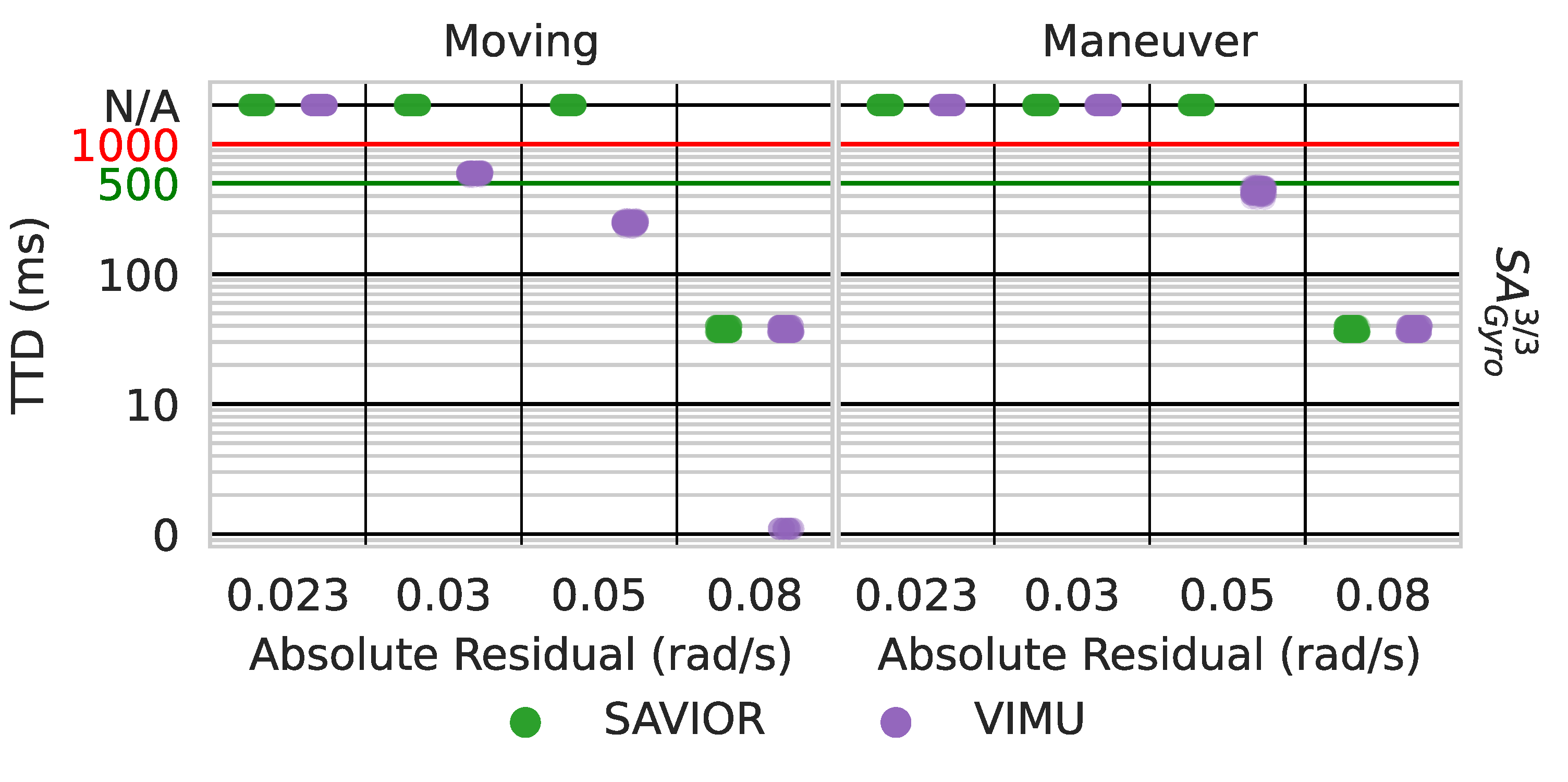}
\vspace{-2ex}
    \caption{TTD on Stealthy Attacks. N/A = No Alarm}
    \label{fig:stealthy_time_to_detect}
\vspace{-2ex}
\end{figure}

Fig.~\ref{fig:time_to_detect} presents the log-scaled TTD of different approaches against overt gyroscope attacks. We omit the TTD on $\OATwoGyro$ because they are similar to $\OAThreeGyro$. \sysname{} has a TTD similar to \savior{} and \vimucusum{} in most attack cases and maintains a steady TTD in small attack deviation, which is favorable in the FIFO buffer design.
Besides, we test with gyroscope attacks at extremely large deviations, as they can disrupt the system state within one sampling interval~\cite{jeong2023unrock}. Based on our quadcopter's sensor specification, we measure the TTD of CS-EMA on $\OAGyroModel{3}{ICM20602}$ and $\OAGyroModel{3}{ICM20689}$ executed at their maximum induced amplitude $A_{max}$ (Section~\ref{subsec:attack_implementation}).
Results show that our CS-EMA detector detects the attacks within one sampling interval (4 ms), thus preventing damage to the flight control.
\revise{In the hovering flights, we observed the same comparative advantages of the CS-EMA detector in TTD on the overt gyroscope attacks.}
Fig.~\ref{fig:gps_joint_attack} shows the TTD under the real-world GPS spoofing $\OAGpsJoint$. \sysname{}, \savior{}, and \vimucusum{} have similar TTDs (7$\sim$9 seconds), indicating that $\OAGpsJoint$ is more complex to detect than the constant-deviation spoofing $\OAGps$ (with TTD $\leq$ 1~second, shown in Fig.~\ref{fig:ttd_gps}).

\subsubsection{Resilience towards Stealthy Attacks}

We evaluate the robustness of detectors by analyzing the maximum $|r^*_i(t)|$ of \eqref{def:stealthy_attack} that the adversary can achieve while remaining stealthy.
A larger $|r^*_i(t)|$ means the adversary can inject more deviation $\Measurement{x}^*_i(t)$ in \eqref{eq:attack_residual} with the same measurement and reference state.
Fig.~\ref{fig:stealthy_time_to_detect} shows three $\SAGps$ and four $\SAGyro$ attacks. These attacks are stealthy because at least one detector under evaluation raises no alarm.
We use TTD to indicate the effectiveness of the alarm.
In Fig.~\ref{fig:stealthy_time_to_detect}, our CS-EMA detector identifies more attacks within the time bound than other detectors.
Moreover, the CS-EMA detector has the tightest bound on stealthy attack deviation.
Both results show that the CS-EMA detector is more robust against stealthy attacks targeting GPS or gyroscope.

\subsection{Recovery Effectiveness}

\begin{figure*}[!tb]
    \centering
    \includegraphics[width=0.98\linewidth]{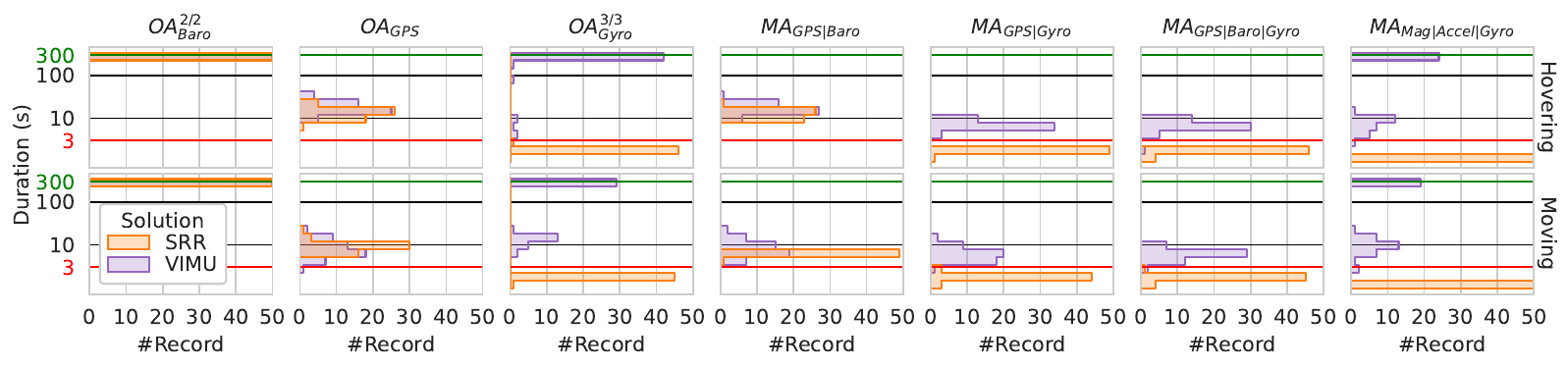}
\vspace{-2ex}
    \caption{Recovery Duration Histogram on Flight Records}
    \label{fig:recovery_duration}
\vspace{-3ex}
\end{figure*}

\subsubsection{Comparison with \SoftwareSensor{}}

We compare \sysname{}'s recovery performance with the state-of-the-art PBAR approach~\SoftwareSensor{}~\cite{choi2020software} in the attacks $\OAGps$, $\OABaro$, $\OAThreeGyro$, and $\MSAttack{*}$.
Both approaches require a detection alarm to initialize the recovery procedure, which means the TTD affects their recovery performance. For a fair comparison, we use the attacks that ensure the immediate initialization of the recovery procedure at the attack-activation moment, e.g., the attack cases of $\OAThreeGyro(0.60)$ with TTD$\approx$0 in Fig.~\ref{fig:time_to_detect}.

Fig.~\ref{fig:recovery_duration} shows the effective recovery duration of \sysname{} and \SoftwareSensor{}.
The durations depend on the type of compromised sensors.
If only the barometers are unavailable ($\OABaro$), the autopilot replaces the altitude source with GPS, and both approaches can achieve a recovery duration close to the upper bound $T_{rec}$.
In contrast, the autopilot of the testbed drone has to rely on the reference states provided by the physical model (or \SoftwareSensor{}'s linear model) in the attack $\OAGps$ and $\OAThreeGyro$ because no other replacement is available for these compromised sensors.
Therefore, the recovery durations in $\OAGps$ and $\OAThreeGyro$ depend heavily on the precision of the physical model.
Specifically, \sysname{} and \SoftwareSensor{} have similar recovery durations in $\OAGps$. The GPS primarily measures the position and velocity states. Both our physical model and \SoftwareSensor{}'s linear model are sufficient to predict their changes over time.
However, the recovery durations diverge in $\OAThreeGyro$ and multi-type sensor attacks involving gyroscopes, because the attitude changes measured by the gyroscope are nonlinear and cannot be accurately predicted by a linear model.
To reduce the error in attitude and angular velocity, \SoftwareSensor{} employs a \emph{supplementary compensation} (SC) mechanism that derives the system attitude from the measurements of the accelerometer and magnetometer~\cite{choi2020software}.
\SoftwareSensor{} activates SC when all gyroscopes are under attack ($\OAThreeGyro$).
Despite this effort, its effective recovery duration in $\OAThreeGyro$ remains less than 3 seconds.
In contrast, \sysname{} succeeds in much longer recovery durations without the SC mechanism, even surpassing the upper bound $T_{rec}$ in many flight records.

\begin{figure}[!tb]
    \centering
    \includegraphics[width=0.9\linewidth]{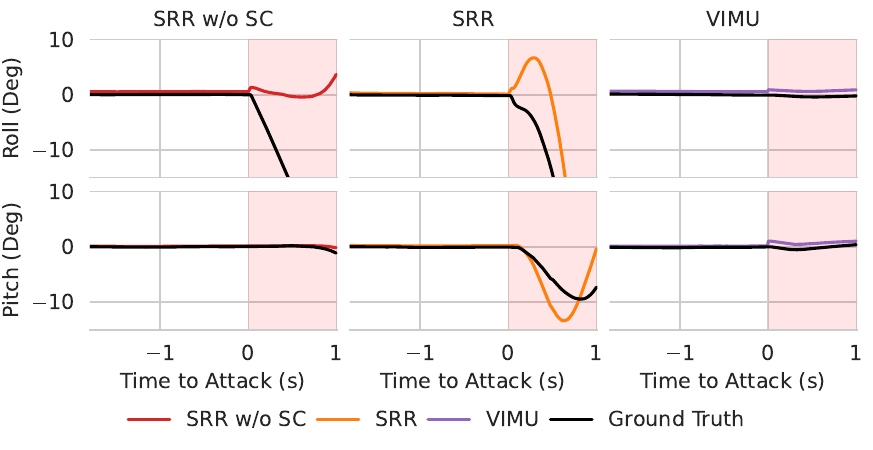}
\vspace{-2ex}
    \caption{Effect of Supplementary Compensation in Attitude Recovery}
    \label{fig:error_in_compensation}
\vspace{-2ex}
\end{figure}

To further investigate the effect of \SoftwareSensor{}'s SC mechanism, we analyze the attitude estimation and its ground truth before and after the activation of $\OAThreeGyro$ attack.
Fig.~\ref{fig:error_in_compensation} presents the roll and pitch angle estimated by \SoftwareSensor{} (with and without SC) and \sysname{}.
The roll estimate of \SoftwareSensor{} without SC diverges immediately since the attack starts (red area), indicating that the linear model of \SoftwareSensor{} is insufficient to capture the non-linearity in attitude changes.
The estimations given by SC also deviate from the ground truth to a certain extent. Such deviation could be due to the acceleration-based estimation requiring several seconds to approach a steady estimate~\cite{leishman2014quadrotors}. In contrast, the attitude estimated by \sysname{} is closer to the ground truth, even without the supplementary compensation.

\subsubsection{Comparison with \savior{} on Physical Model}

\begin{figure}[!tb]
    \centering
    \includegraphics[width=0.9\linewidth]{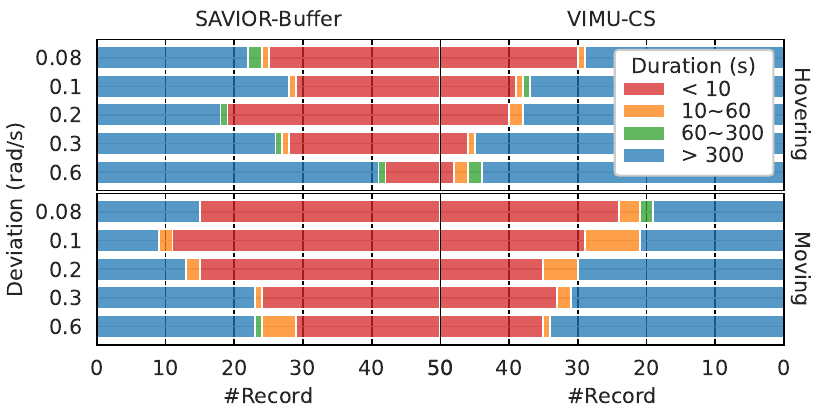}
\vspace{-2ex}
    \caption{Recovery Durations (\vimucusum{} vs. \saviorbuffer{})}
    \label{fig:recovery_duration_with_savior}
\vspace{-3ex}
\end{figure}

Our physical model retrofits \savior{}'s nonlinear physical model~\cite{quinonez2020savior} with a more accurate specification in estimating motor thrust and aerodynamic drag on the airframe. Fig.~\ref{fig:roc_curves} has depicted the physical model's contribution to the detection effectiveness.
Here, we further investigate its contribution to recovery effectiveness.
Considering the effect of the FIFO buffer, we compare \saviorbuffer{} with \vimucusum{} regarding the recovery duration in attack $\OAThreeGyro$.
Fig.~\ref{fig:recovery_duration_with_savior} presents the results.
We exclude the deviation settings at 0.04 and 0.06 rad/s because neither approach can detect these attacks to launch the recovery.
For the rest deviation settings, we observed that \vimucusum{} outperforms \saviorbuffer{} in the average recovery duration by 8\%$\sim$123\%, w.r.t.
different attack deviations.
Since \saviorbuffer{} and \vimucusum{} have the same detector (and parameters) and FIFO buffers, the differences in recovery duration reflect the discrepancy between the physical models.

\subsubsection{Real World Test}\label{subsubsec:realworld}

We demonstrate \sysname{}'s recovery effect on the real-world quadcopter by injecting an $\OAThreeGyro$ attack.
In the test, the drone took off from home and hovered at the preset altitude. Then, we launched the $\OAThreeGyro$ attack on all three gyroscopes by injecting constant deviation (0.60 rad/s) to disrupt the gyroscopes' roll rate measurements. The drone with \sysname{}'s detection and recovery prevented the immediate crash and maintained the flight attitude for around 26.1 seconds\footnote{\url{https://www.youtube.com/watch?v=IVZsC0wPPA8}}.
As a comparison, we conducted another flight with \sysname{} but disabled \sysname{}'s recovery. Without the recovery, the drone immediately deviated from its hovering position and crashed\footnote{\url{https://www.youtube.com/watch?v=WAuYQXz_tIE}}.

\revise{
\emph{Comparison on TTD}. The real-world IMU data quality causes longer TTD in attack detection as the detector needs more sampling periods to distinguish the attack from sensor noises. To investigate CS-EMA's advantage over other detectors, e.g., CUSUM, we compare \sysname{} with \vimucusum{} in TTD with the attack $\OAThreeGyro$ to real-world flights. We did not use \savior{} because we cannot properly tune \savior{}'s physical model on our quadcopter. The results show that the CS-EMA detector takes 19.4$\sim$20.7 ms to raise the alarm, while the CUSUM detector fails to identify the attack in 1 second, i.e., $T^\text{Gyro}_\text{Alarm}$. Moreover, in the same mission flight without applying attack, CS-EMA reports no false positive.
}

\revise{
\emph{Energy Consumption and Runtime Overhead}.
We perform real flights with our quadcopter. We use the \texttt{battery\_status} message to evaluate the power consumption during the flight. The gross power of our drone is around 350 watts, mostly ($>$340 watts) consumed by the actuators.
The onboard ARM CPU consumes 0.85 watts at 100\% usage.
We then evaluate the performance overhead of \sysname{} with onboard CPU and memory usage. We use the \texttt{cpuload} message to record the runtime overhead at the armed, hover, and moving conditions.
Table~\ref{table:runtime_overhead} presents the runtime overhead of the autopilot with and without the \sysname{} deployment.
On average of the flight maneuvers, the CPU usage has increased by 15.65\% (peak) and 16.74\% (mean), and the overall memory has increased by 54 KB.
As a result, the energy overhead of \sysname{} is around 0.15 watts (0.043\%), which is negligible compared with the total power consumption.
}

\begin{table}[tb]
\renewcommand{\arraystretch}{1.2}
\caption{Runtime Overheads with Onboard CPU}
\label{table:runtime_overhead}
\vspace{-1ex}
\centering
\resizebox{\columnwidth}{!}{
    \begin{tabular}{c|c|c|c|c|c|c|c}
        \hline
        \multicolumn{2}{l|}{} & \multicolumn{2}{c|}{Armed} & \multicolumn{2}{c|}{Hover} & \multicolumn{2}{c}{Moving} \\
        \cline{3-8}
        \multicolumn{2}{l|}{} & PX4 & VIMU & PX4 & VIMU & PX4 & VIMU \\
        \hline
        \multirow{2}*{CPU (\%)} & Mean & 51.34 & 65.24 & 50.94 & 69.77 & 51.61 & 69.10 \\
        \cline{2-8}
         & Peak & 57.00 & 68.54 & 52.46 & 70.67 & 52.65 & 69.84 \\
         \hline
        RAM (\%) & Peak & 57.73 & 67.68 & 71.96 & 82.66 & 71.96 & 82.66 \\
        \hline
    \end{tabular}
}
\vspace{-3ex}
\end{table}

\section{Discussion}
\label{sec:discussion}

Our CS-EMA detector alerts sensor attacks faster and has a tight TTD.
Although our evaluation mainly focuses on spoofing attacks, these results can generalize to sensor attacks that aim for a denial of service, e.g.,~\cite{jang2023paralyzing}. These attacks will introduce large deviations to the sensor measurements, which makes them easily identified by our detector (Section~\ref{subsec:effectiveness_on_overt_attacks}).
Besides, such an advantage also benefits the development of the FIFO buffer (Section~\ref{subsec:method_ekf}), as TTD positively correlates with $\TBuffer$ in \eqref{def:buf_size}.
By the definition of $size_\text{buffer}$ in \eqref{def:buf_size}, lower TTD results in a reduced memory cost to implement the buffers.
Given the results in Fig.~\ref{fig:time_to_detect}, we set $\TBuffer = 500$ ms for \sysname{}, which is sufficient to cover the TTD in most cases.
For the related PBAR approach \SoftwareSensor{}~\cite{choi2020software}, a much larger FIFO buffer is needed (4x to 10x per IMU compared with \sysname{}) because \SoftwareSensor{}'s detector encounters numerous ineffective alarms (TTD$~\geq T_\text{Alarm}$).
For \saviorbuffer{} used in $\OAThreeGyro(\ge 0.08)$, its FIFO buffer size can generally equal to \sysname{}'s buffer size, since the detectors of \savior{} and \sysname{} have their respective merit on these attack cases in Fig.~\ref{fig:time_to_detect}.
We have examined the FIFO buffer's contribution to the recovery duration. After using the buffer safeguard, the average recovery duration increases by 66\%$\sim$746\%, w.r.t. different TTDs (Appendix~\ref{app:detection_effectiveness}).
Moreover, the effectiveness of EMA as a low-pass filter offers the potential for applying other forms of low-pass filter, e.g., wavelet transform~\cite{DBLP:journals/compsec/CallegariGPP12}, to our detection scenario to distinguish persistent deviation from the high-frequency measurement noise.

\section{Conclusion}

As an effective and efficient PBAR approach against sensor attacks, \sysname{} uses an accurate nonlinear physical model to estimate the system states for the sensor-compromised autopilot. \sysname{}'s anomaly detector deploys a CS-EMA detector to improve detection effectiveness and reduce the detection time delay on high-rate gyroscope attacks.
With a properly learned physical model and selected detector parameters, \sysname{} outperforms the state-of-the-art PBAD and PBAR approaches regarding the effectiveness and efficiency of detection and system-state recovery.




\bibliographystyle{IEEEtranS}
\bibliography{IEEEabrv,refs}

\appendices

\section{Physical Parameters Determination}\label{app:Model_Learning}

The physical model described in Section~\ref{subsec:physical model} applies to any commercial quadcopter.
Although the parameters of the physical model may vary by the airframe specification, the learning process usually only needs once for each airframe type, and the specific parameters will be deployed as default drone product settings.
This section determines the physical parameters $\ModelParams$ in \eqref{def:model_params}. Specifically, we divide $\ModelParams$ into $\ModelParams_\text{measure}=\{ \VehicleMass, \ArmLength, \InertiaMatrix, \ThrustCoeff, \TorqueCoeff, \ThrustSpZero, \ThrustSpRange, \VoltageRef, \ResistanceInternal\}$ and $\ModelParams_\text{learn}=\{\TorqueRotorCoeff, \TimeConstant, \MCoef, \BCoefVec\}$.

\emph{Determination of $\ModelParams_\text{measure}$}.
The drone mass $\VehicleMass$ and the distance $\ArmLength$ between the motor and the airframe's CoG are directly measured on the specific drone airframe.
We determine the diagonal inertia terms $I_{xx}$, $I_{yy}$, and $I_{zz}$ of the inertia matrix $\InertiaMatrix$ through torsional pendulum tests \cite{setati2022experimental}.
We omit the non-diagonal elements in $\InertiaMatrix$ by assuming the airframe is symmetrical~\cite{quinonez2020savior, tomic2020simultaneous}.
To ensure the model precision, we measure $\ThrustCoeff$, $\TorqueCoeff$, and $\ThrustSpZero$ with a static test stand\footnote{\url{https://www.youtube.com/watch?v=O6upmU9ubPQ}}.
Our drone uses PWM-based motor controllers, which give us $\ThrustSpRange = 1000$ us.
We read $\ResistanceInternal$ from a RadioLink CB86-PLUS balance charger.
For the 6-cell LiPo battery used by our drone, the full capacity voltage of each cell is $4.05$ V, and we have $\VoltageRef = 24.3$ V.

\emph{Determination of $\ModelParams_\text{learn}$}.
Based on \eqref{def:fitting}, we fit the physical parameters with known system states $\ModelState(t)$ and control inputs $\ModelInput(t)$ from the normal flight records.
We learn $\TorqueRotorCoeff$, $\TimeConstant$, $\MCoef$, $\BCoefX$ and $\BCoefY$ of $\BCoefVec$ through the Nelder-Mead \cite{wright1996direct} searching over the nonlinear least squares data fitting problem~\cite{griva2009linear}.

Given the input data $X$, the fitting target $Y$, the differential equations $F(\cdot)$ describing the physical model, and the parameter set $\ModelParams$, the nonlinear least squares method decides the optimal parameters by minimizing the squared error:
\begin{align}
    \label{eq:estimation_error}
    \min _{\mathcal{P}} \sum_{t=1}^T\left(F_t\left(X_t; \ModelParams \right)-Y_t\right)^2
\end{align}
where $F_t\left(X_t; \ModelParams \right)$ is the state at time $t$ predicted by $F(\cdot)$. $Y_t$ is the estimated state or sensor measurement from the flight record w.r.t. the prediction result $F_t\left(X_t; \ModelParams \right)$.

Before deciding the specific parameters of $\ModelParams$, we first follow the guides\footnote{\url{https://docs.px4.io/main/en/advanced_config/tuning_the_ecl_ekf.html}}$^,$\footnote{\url{https://docs.px4.io/main/en/config/autotune.html}}
to obtain the following data:
1) sensor measurements $\Measurement{\rho}$, $\Measurement{\AngRateVec}$, and body acceleration $\Measurement{\VecStyle{a}}^B$, 2) estimated states $\Estimate{\quaternion}$, $\Estimate{\vned}$, and $\EstimateWindSpeed{}$, 3) $\ModelInput(t)$ and 4) $\VoltageLoad$ and $\CurrentLoad$.
This data collection process is conducted for one time on the standard instance of the airframe model.
Here, we used the quaternion form of system attitude, i.e., $\quaternion = [q_w, q_i, q_j, q_k]^T$, as the substitute of the rotation matrix $\dcm$ in Section \ref{subsec:physical model}.
The conversion between the quaternion and $\dcm$ is straightforward~\cite{shuster1993survey}.
Using $\quaternion$ instead of $\dcm$ can reduce the computational cost of learning because it is commonly used by the autopilots, e.g., PX4 and ArduPilot.

\begin{table}[!tb]
\renewcommand{\arraystretch}{1.2}
  \caption{Physical Parameters in Simulation and Real-world}
  \label{tab:physical-parameters-simulation}
  \label{tab:physical-parameters-realworld}
  \centering
\vspace{-2ex}
\resizebox{\columnwidth}{!}{
  \begin{tabular}{c|c|c|l}
    \hline
    Param             & \sysname{} (Simulation)  & \sysname{} (Real-world)                       & Unit    \\ \hline
    $\VehicleMass$        &  0.80                   &  2.64                       & kg    \\
    $\ArmLength$          &  0.165                   &  0.288                      & m   \\
    $\InertiaMatrix$      &  diag\{5.0, 5.0, 9.0\} $\cdot 10^{-3}$ &  diag\{5.17, 5.50, 7.62\} $\cdot 10^{-2}$ & kg$\cdot$m$^2$   \\
    $\ThrustCoeff$        &  4.0                     &  23.0                       & N  \\
    $\TorqueCoeff$        &  0.05                    &  0.44                       & N$\cdot$m   \\
    $\TorqueRotorCoeff$   &  0.0                     &  0.061                      & N$\cdot$m/s    \\
    $\TimeConstant{}$     &  0.005                            &  0.035                      & s    \\
    $\ThrustSpZero$       &  1000                  &  840                        & us    \\
    $\ThrustSpRange$      &  1000                    &  1000                        & us    \\
    $\VoltageRef$         &  N/A       &  24.3                       & V     \\
    $\ResistanceInternal$ &   N/A       &  0.072                      & $\Omega$    \\
    $\MCoef$              &  0.001                   &  0.031                      & 1/s    \\
    $\BCoefVec$           &  [0.022, 0.022, 0]       &  [0.161, 0.145, 0]          & m$^2$/kg   \\
  \hline
\end{tabular}
}
\vspace{-4ex}
\end{table}

According to \cite{leishman2014quadrotors}, the accelerometer measures the specific acceleration $\Measurement{\VecStyle{a}}^B = [\Measurement{a}_x, \Measurement{a}_y, \Measurement{a}_z]^T$. Indeed, $\Measurement{\VecStyle{a}}^B$ stands for the difference between the vehicle's acceleration ($\Aned$ in \eqref{eq:acc_change}) and gravitational acceleration ($\Gee \unitvec$).
This measurement is already located in the FRD frame and thus requires no rotation to align with $\CtrlAccel$ and $\DragAccel$. From that, we have $\Measurement{\VecStyle{a}}^B = \CtrlAccel + \DragAccel$.
Since $\CtrlAccel$ is generally vertical to the x-axis and y-axis, $\Measurement{a}_x$ and $\Measurement{a}_y$ are dominated by the drag acceleration $\DragAccel$.
Therefore, we use $\Measurement{a}_x$ and $\Measurement{a}_y$ to solve $\MCoef$, $\BCoefX$, and $\BCoefY$ with \eqref{eq:drag_accel}.
We first derive the relative airspeed $\Airspeed$ from the estimated $\Estimate{\quaternion}$, $\Estimate{\vned}$, and $\EstimateWindSpeed{}$. Then, we calculate the vector norm $\lVert \Airspeed \rVert$ in \eqref{eq:drag_accel}.
After that, we replace $\DragAccel$ with $[\Measurement{a}_x, \Measurement{a}_y]$ of $\Measurement{\VecStyle{a}}^B$ and use $\lVert \Airspeed \rVert$, $\Measurement{\rho}$, and the x-axis and y-axis components of $\Airspeed$ to solve the equation.
In short, we consider $X_1 = \{ \Estimate{\quaternion}, \Estimate{\vned}, \EstimateWindSpeed{}, \Measurement{\rho} \}$ as input data and $Y_1 = \{ \Measurement{a}_x, \Measurement{a}_y \}$ as the fitting target in \eqref{eq:drag_accel} to resolve $\MCoef$, $\BCoefX$, and $\BCoefY$.
We compute the estimate of $\DragAccel$ and use it to derive $\CtrlAccel = \Measurement{\VecStyle{a}}^B - \DragAccel$.
Then, we solve \eqref{eq:ctrl_accel}, \eqref{eq:thrust_adjust}, and \eqref{eq:thrust_delay} to obtain $\TimeConstant$ with $X_2 = \{\ModelInput(t), \VoltageLoad, \CurrentLoad\}$ as input data and $Y_2 = \CtrlAccel$ as the fitting target.
Finally, we solve $\TorqueRotorCoeff$ in \eqref{eq:ctrl_torque} with the input data $X_3=\{\ModelInput(t), \Measurement{\AngRateVec}, \VoltageLoad, \CurrentLoad\}$ and the fitting target $Y_3 = \AngAccVec(t)$, where $\AngAccVec(t)$ is estimated from the angular velocity $\Measurement{\AngRateVec}$ at timestamp $t$ and $t-1$.

\section{Detector Parameter Selection}
\label{app:Param_Selection}

We select the most suitable detector parameters based on two criteria: 1) the FPR in the survey mission of QGroundControl \revise{or real flight}, and 2) the theoretical TTD at possible attack deviations.
The survey mission contains multiple waypoints in its task route.
It covers flight maneuvers that any drone will encounter in common flight scenarios, e.g., takeoff, hover, cruise, and turn.
The diversity in flight maneuvers qualifies the survey mission for parameter selection, as the detector performance in attack-free flight is mainly affected by the measurement noise and modeling error.

\begin{figure}[tb!]
    \centering
    \includegraphics[width=0.95\linewidth]{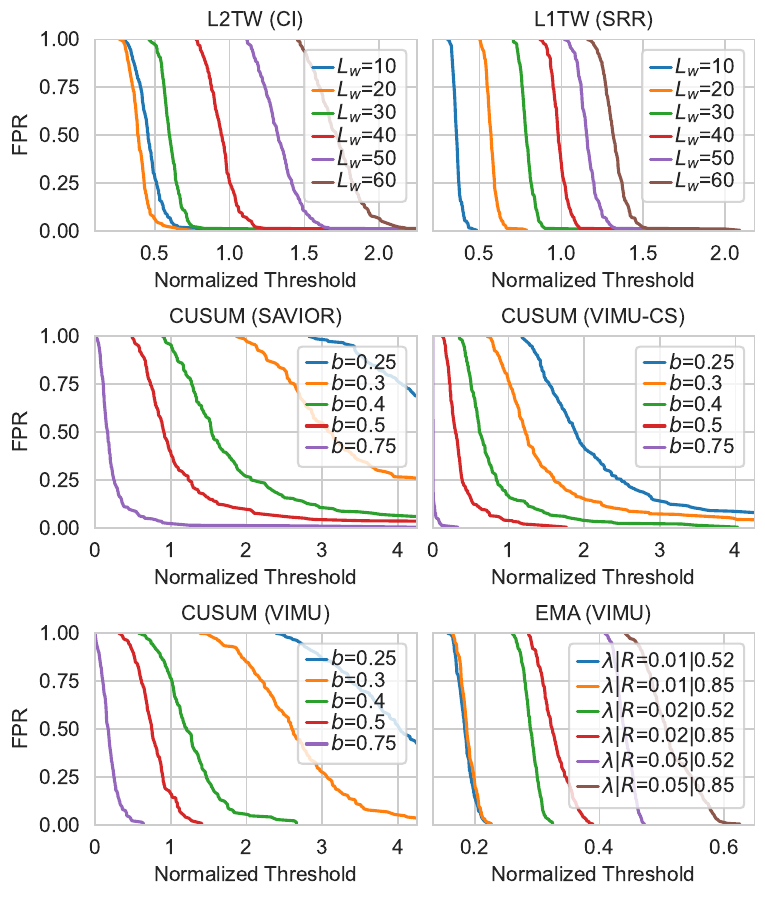}
\vspace{-2ex}
    \caption{FPR-Threshold Curves of Gyroscope Detectors at Different Parameters}
    \label{fig:param_selection_default}
\vspace{-2ex}
\end{figure}

We first collect the flight data for parameter selection.
For each approach, we use the survey mission to collect 100 attack-free flight records. During this process, we turn off the isolation and recovery functionality to ensure the autopilot can complete the mission.
\revise{
For each sensor instance, we calculate the in-flight residuals to construct the validation dataset.
Then, we determine the candidate detector parameter settings (e.g., different value pairs of $(\lambda, R)$ for \sysname{}'s EMA detector) and use the validation dataset to decide their corresponding detection threshold (e.g., $\tau_{ema}$ for \sysname{}'s EMA detector). The specific steps are as follows.}
\revise{
\begin{compactenum}[1)]
  \item For each setting of candidate detector parameters, we use the validation set of each flight to figure out the \emph{minimum false-alarm-avoidance threshold} ($\tau_\textit{avoid}$) of the flight. We rank the $\tau_\textit{avoid}$ of the 100 flights in ascending order to obtain one curve in a subfigure of Fig.~\ref{fig:param_selection_default}. Due to the ascending order of $\tau_\textit{avoid}$, the vertical axes of Fig.~\ref{fig:param_selection_default} represent the FPR of the gyroscope detector configured with one flight's $\tau_\textit{avoid}$ and the specific candidate detector parameters when applied on other flights. We call this curve \emph{FPR-Threshold curve} w.r.t. specific candidate detector parameters.
  \item We define the fifth largest $\tau_\textit{avoid}$ on each FPR-Threshold curve as the \emph{reference detection threshold} ($\tau_\textit{ref}$) of each candidate detector parameter setting.
      In other words, the detector equipped with $\tau_\textit{ref}$ should cause no more than five false alarms in the 100 missions.
  \item We take different candidate detector parameter settings and the corresponding $\tau_\textit{ref}$ to configure the gyroscope detector and figure out the TTDs under these detector configurations at various attack deviations. Then, we confirm the most appropriate detector parameter setting (e.g., $(\lambda, R)=(0.01,0.85)$ for the EMA detector) and the corresponding $\tau_\textit{ref}$. We add a \emph{safety margin} to this $\tau_\textit{ref}$ to obtain the final detection threshold $\tau$. Specifically, $\tau =1.05\cdot \tau_\textit{ref}$.
\end{compactenum}
}
\noindent Following the above procedure, Table~\ref{table:baseline_detector_params} presents the detector parameters used in the experiments in Section~\ref{sec:evaluation}.
Note that, to conform with the autopilot implementation, we scale the residual defined in \eqref{def:residual}, detection thresholds $\tau$, mean shifts $b_i$, and $R$ to the standard deviation by dividing them by the noise parameters.

\begin{table}[!tb]
\renewcommand{\arraystretch}{1.2}
\caption{Detector Parameters}
\label{table:baseline_detector_params}
\label{table:our_detector_params}
\centering
\vspace{-2ex}
\resizebox{\columnwidth}{!}{
\begin{tabular}{c|cc|cc|cc|ccccc}
\hline
Detector & \multicolumn{2}{c|}{L1TW} & \multicolumn{2}{c|}{L2TW} & \multicolumn{2}{c|}{CUSUM} & \multicolumn{5}{c}{CS-EMA} \\
\hline
Measurement & \multicolumn{1}{c|}{$\tau$} & $L_w$ & \multicolumn{1}{c|}{$\tau$} & $L_w$ & \multicolumn{1}{c|}{$\tau$} & $b_i$ & \multicolumn{1}{c|}{$\tau_{cs}$} & \multicolumn{1}{c|}{$b_i$} & \multicolumn{1}{c|}{$\tau_{ema}$} & \multicolumn{1}{c|}{$\lambda$} & $R$ \\
\hline
GPS Position & \multicolumn{1}{c|}{1.15} & 10 & \multicolumn{1}{c|}{1.402} & 10 & \multicolumn{1}{c|}{3.0} & 0.50 & \multicolumn{1}{c|}{3.0} & \multicolumn{1}{c|}{0.50} & \multicolumn{1}{c|}{0.45} & \multicolumn{1}{c|}{0.01} & 0.85 \\
GPS Velocity & \multicolumn{1}{c|}{3.10} & 10 & \multicolumn{1}{c|}{4.42} & 10 & \multicolumn{1}{c|}{3.5} & 1.00 & \multicolumn{1}{c|}{3.5} & \multicolumn{1}{c|}{1.00} & \multicolumn{1}{c|}{0.50} & \multicolumn{1}{c|}{0.01} & 1.10 \\
Barometer & \multicolumn{1}{c|}{0.10} & 10 & \multicolumn{1}{c|}{0.02} & 10 & \multicolumn{1}{c|}{3.0} & 0.25 & \multicolumn{1}{c|}{3.0} & \multicolumn{1}{c|}{0.25} & \multicolumn{1}{c|}{0.15} & \multicolumn{1}{c|}{0.05} & 0.52 \\
Magnetometer & \multicolumn{1}{c|}{0.35} & 10 & \multicolumn{1}{c|}{0.2} & 10 & \multicolumn{1}{c|}{3.0} & 0.25 & \multicolumn{1}{c|}{3.0} & \multicolumn{1}{c|}{0.25} & \multicolumn{1}{c|}{0.30} & \multicolumn{1}{c|}{0.01} & 0.52 \\
Gyroscope & \multicolumn{1}{c|}{0.464} & 10 & \multicolumn{1}{c|}{0.65} & 20 & \multicolumn{1}{c|}{3.0} & 0.50 & \multicolumn{1}{c|}{3.0} & \multicolumn{1}{c|}{0.50} & \multicolumn{1}{c|}{0.25} & \multicolumn{1}{c|}{0.01} & 0.85 \\
Accelerometer & \multicolumn{1}{c|}{5.20} & 10 & \multicolumn{1}{c|}{30.25} & 10 & \multicolumn{1}{c|}{3.0} & 1.00 & \multicolumn{1}{c|}{3.0} & \multicolumn{1}{c|}{1.00} & \multicolumn{1}{c|}{0.95} & \multicolumn{1}{c|}{0.01} & 1.10 \\
\hline
\end{tabular}
}
\vspace{-2ex}
\end{table}

\begin{table}[!tb]
\renewcommand{\arraystretch}{1.2}
\caption{Impact of FPR Standard on Parameter Selection}
\vspace{-2ex}
\label{table:false_positive_with_savior}
\resizebox{\columnwidth}{!}{
\begin{tabular}{c|ccc|ccc}
\hline
$\tau$ (Actual FPR &
  \multicolumn{3}{c|}{CUSUM (SAVIOR)} &
  \multicolumn{3}{c}{CUSUM (VIMU)} \\ \cline{2-7}
under $\tau$) &
  \multicolumn{1}{c|}{1\%} &
  \multicolumn{1}{c|}{3\%} &
  5\% &
  \multicolumn{1}{c|}{1\%} &
  \multicolumn{1}{c|}{3\%} &
  5\% \\
\hline
b = 0.25 &
  \multicolumn{1}{c|}{N/A} &
  \multicolumn{1}{c|}{N/A} &
  N/A &
  \multicolumn{1}{c|}{N/A} &
  \multicolumn{1}{c|}{N/A} &
  N/A \\ \hline
b = 0.3 &
  \multicolumn{1}{c|}{N/A} &
  \multicolumn{1}{c|}{N/A} &
  N/A &
  \multicolumn{1}{c|}{4.5 (2\%)} &
  \multicolumn{1}{c|}{4.5 (2\%)} &
  4.5 (2\%) \\ \hline
b = 0.4 &
  \multicolumn{1}{c|}{N/A} &
  \multicolumn{1}{c|}{N/A} &
  5.5 (5\%) &
  \multicolumn{1}{c|}{3.0 (0\%)} &
  \multicolumn{1}{c|}{3.0 (0\%)} &
  3.0 (0\%) \\ \hline
b = 0.5 &
  \multicolumn{1}{c|}{N/A} &
  \multicolumn{1}{c|}{5.5 (3\%)} &
  3.0 (4\%) &
  \multicolumn{1}{c|}{3.0 (0\%)} &
  \multicolumn{1}{c|}{3.0 (0\%)} &
  3.0 (0\%) \\ \hline
b = 0.75 &
  \multicolumn{1}{c|}{3.0 (1\%)} &
  \multicolumn{1}{c|}{3.0 (1\%)} &
  3.0 (1\%) &
  \multicolumn{1}{c|}{3.0 (0\%)} &
\multicolumn{1}{c|}{3.0 (0\%)} &
  3.0 (0\%) \\ \hline
\end{tabular}
}
\vspace{-4ex}
\end{table}

\revise{\emph{Rationality of 5\%-FPR standard to decide $\tau_\textit{ref}$}.
The safety margin added onto the reference threshold $\tau_\textit{ref}$ ensures that the final detection threshold $\tau$ has an FPR much lower than $\tau_\textit{ref}$.
Since the detector's sensitivity could be affected by this FPR standard used to decide $\tau_\textit{ref}$, we discuss how the variation of this FPR standard can affect the final $\tau$.
Taking the CUSUM gyroscope detectors as an example, Table~\ref{table:false_positive_with_savior} presents the final detection thresholds $\tau$ decided by different FPR standards (1\%, 3\%, and 5\%).
N/A means the obtained threshold is ineffective in identifying overt attacks, thus not applicable for attack detection.
Compared with \savior{}, most thresholds of \sysname{}'s CUSUM component can achieve an actual FPR of 0.}

\begin{figure*}[tb!]
    \centering
    \includegraphics[width=0.95\linewidth]{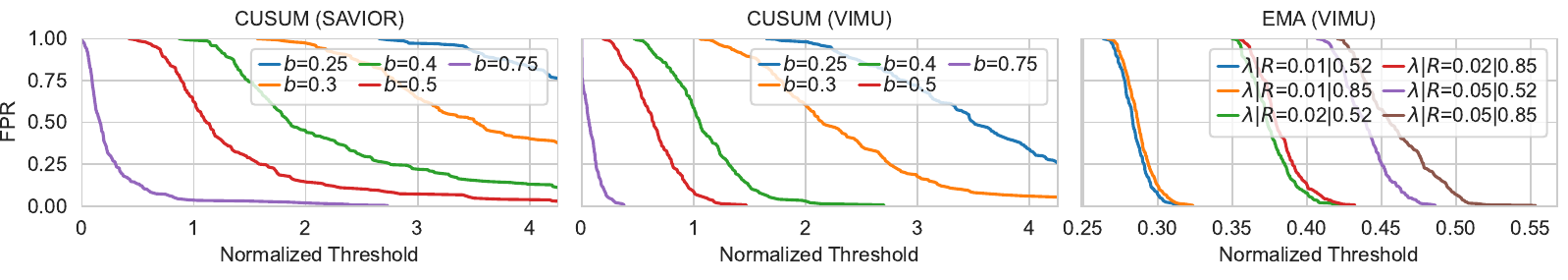}
\vspace{-2ex}
    \caption{FPR-Threshold Curves of Gyroscope Detectors at
Different Parameters in \textit{Maneuver} Mission}
    \label{fig:parameter_selection_maneuver}
\vspace{-3ex}
\end{figure*}

\revise{
\emph{Deciding Detector Parameters for \textit{Maneuver} Scenario and Real-World Drone}.
The \textit{Maneuver} scenario involves several drastic attitude changes. Therefore, we estimate the impact on detector parameter selection.
We confirmed that only the parameters of gyroscope detectors need to be reconfigured.
Compared with Fig.~\ref{fig:param_selection_default}, the FPR-Threshold curves in Fig.~\ref{fig:parameter_selection_maneuver} show higher FPR at the same threshold, indicating that drastic maneuvers cause more false alarms during the attack-free flight.
Meanwhile, Fig.~\ref{fig:parameter_selection_maneuver} shows \sysname{}'s CUSUM component keeps an FPR lower than \savior{} at the same detector parameters ($\tau_\textit{avoid}$, $b_i$), which means our solution has fewer false alarms than the \savior{}.
Following the same selection procedure, we adjust the $\tau_{ema}$ of gyroscope detectors from 0.25 to 0.32. The detection effectiveness in this scenario is presented in Appendix~\ref{app:evaluation_maneuver}.
On the other hand, we take this procedure to decide the detector parameters for the real-world quadcopter.
We fly the quadcopter to collect 100 attack-free flight logs and use these logs to decide the detector parameter. For the gyroscope detector, the specified parameters are $\tau_{cs} = 3.0$, $b_i = 0.75$, $\tau_{ema} = 0.22$, $R = 0.52$, and $\lambda = 0.075$.
Such parameters of the CS-EMA and CUSUM detectors are deployed on our quadcopter to obtain the real-world TTD evaluation results in Section~\ref{subsubsec:realworld}.
}

\begin{figure}[tb!]
    \centering
    \includegraphics[width=0.85\linewidth]{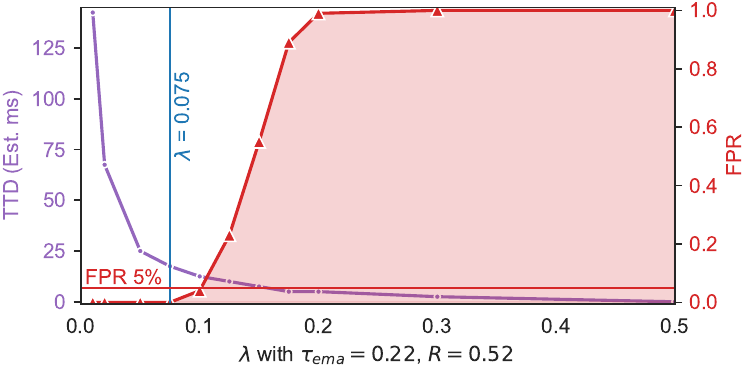}
\vspace{-2ex}
    \caption{$\lambda$'s Sensitivity in Real-World}
    \label{fig:lambda_sensitivity}
\vspace{-2ex}
\end{figure}

\revise{\emph{Sensitivity of $\lambda$}.
The value of $\lambda$ in \eqref{def:stat_ema_moving_average} affects the detector performance.
We discuss the sensitivity of $\lambda$ with real-world flight.
With fixed $\tau_{ema}$ and $R$, Fig.~\ref{fig:lambda_sensitivity} presents the real-world impact of $\lambda$ selection on the FPR and the TTD of $\OAThreeGyro(0.60)$. It indicates that the $\lambda$ selection is a trade-off between TTD and FPR. We also observed that a larger $\tau_{ema}$ results in a lower FPR and a longer TTD. Note that in Fig.~\ref{fig:lambda_sensitivity}, both the FPR and TTD remain relatively stable when $\lambda \in [0.05, 0.10]$, which means in this range of $\lambda$, the detector performance is less sensitive to $\lambda$, justifying our choice of $\lambda=0.075$.
}

\section{Evaluation Results Supplementary}\label{app:evaluation-supplementary}

\subsection{Detection and Recovery Effectiveness}\label{app:detection_effectiveness}

\begin{figure}[tb!]
    \centering
    \includegraphics[width=0.9\linewidth]{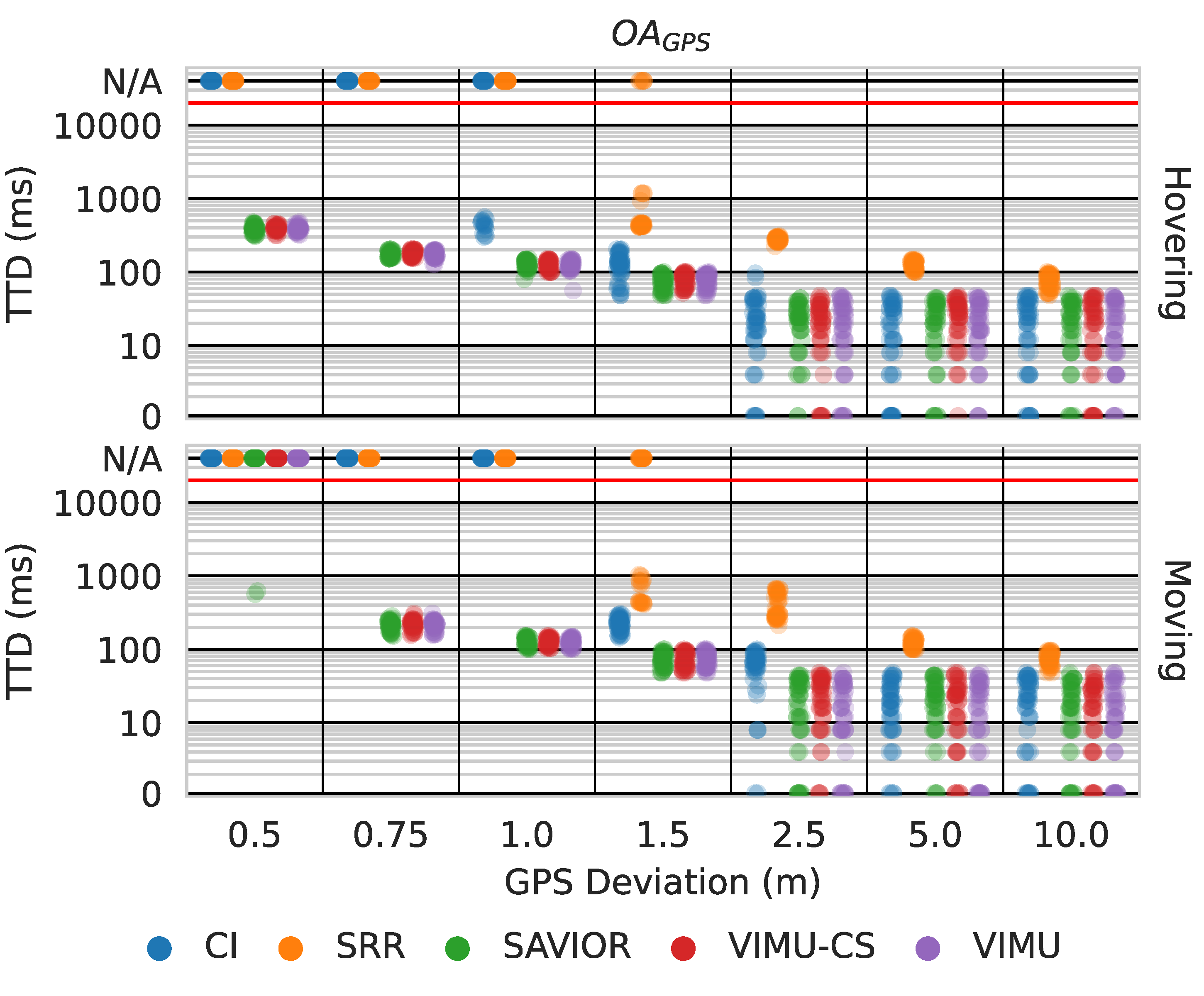}
\vspace{-2ex}
    \caption{Time to Detect on $\OAGps$ (Red Line: $T^\text{GPS}_\text{Alarm}$)}
    \label{fig:ttd_gps}
\vspace{-1ex}
\end{figure}

Fig.~\ref{fig:ttd_gps} presents the TTD under $\OAGps$.
\sysname{}, \savior, and \vimucusum{} have similar TTD performance on this attack. The TTDs are generally less than one second; thus, detecting $\OAGps$ is much simpler than detecting $\OAGpsJoint$, as can be compared with Fig.~\ref{fig:gps_joint_attack}.

\begin{figure}[!tb]
    \centering
    \includegraphics[width=0.88\linewidth]{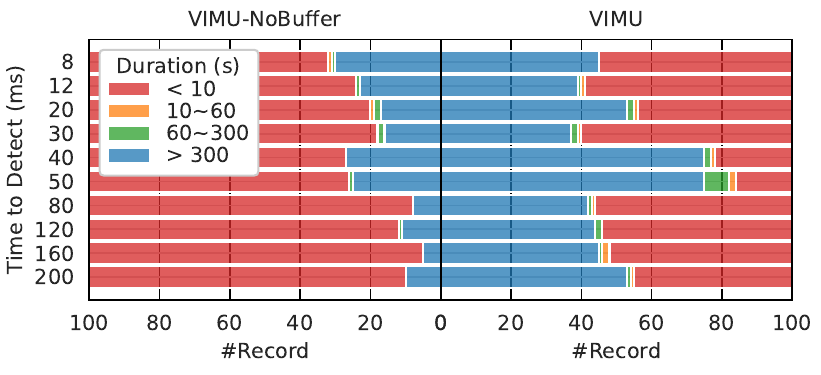}
\vspace{-2ex}
    \caption{Recovery Duration of \sysname{} in Different TTDs}
    \label{fig:effect_of_buffer}
\vspace{-2ex}
\end{figure}

We also evaluate the impact of TTD and FIFO buffer on recovery duration with attack $\textit{Hovering}[\OAThreeGyro(0.60)]$.
In Fig.~\ref{fig:effect_of_buffer}, the recovery duration of \sysname{} without the FIFO buffer generally reduces as the TTD increases, indicating that the adversary can cause more damage to the system state in the detections with longer TTD.
When evaluating the complete \sysname{} (with the FIFO buffer), we observed that the average recovery duration increases by 66\%$\sim$746\%, w.r.t. different TTDs. The distribution of recovery duration becomes insensitive to TTD with the buffer deployed.
These results showed that the FIFO buffer effectively mitigates high-rate sensor attacks.

\subsection{Impact of Drastic Flight Maneuvers}
\label{app:evaluation_maneuver}

\begin{figure}[tb!]
    \centering
    \includegraphics[width=0.9\linewidth]{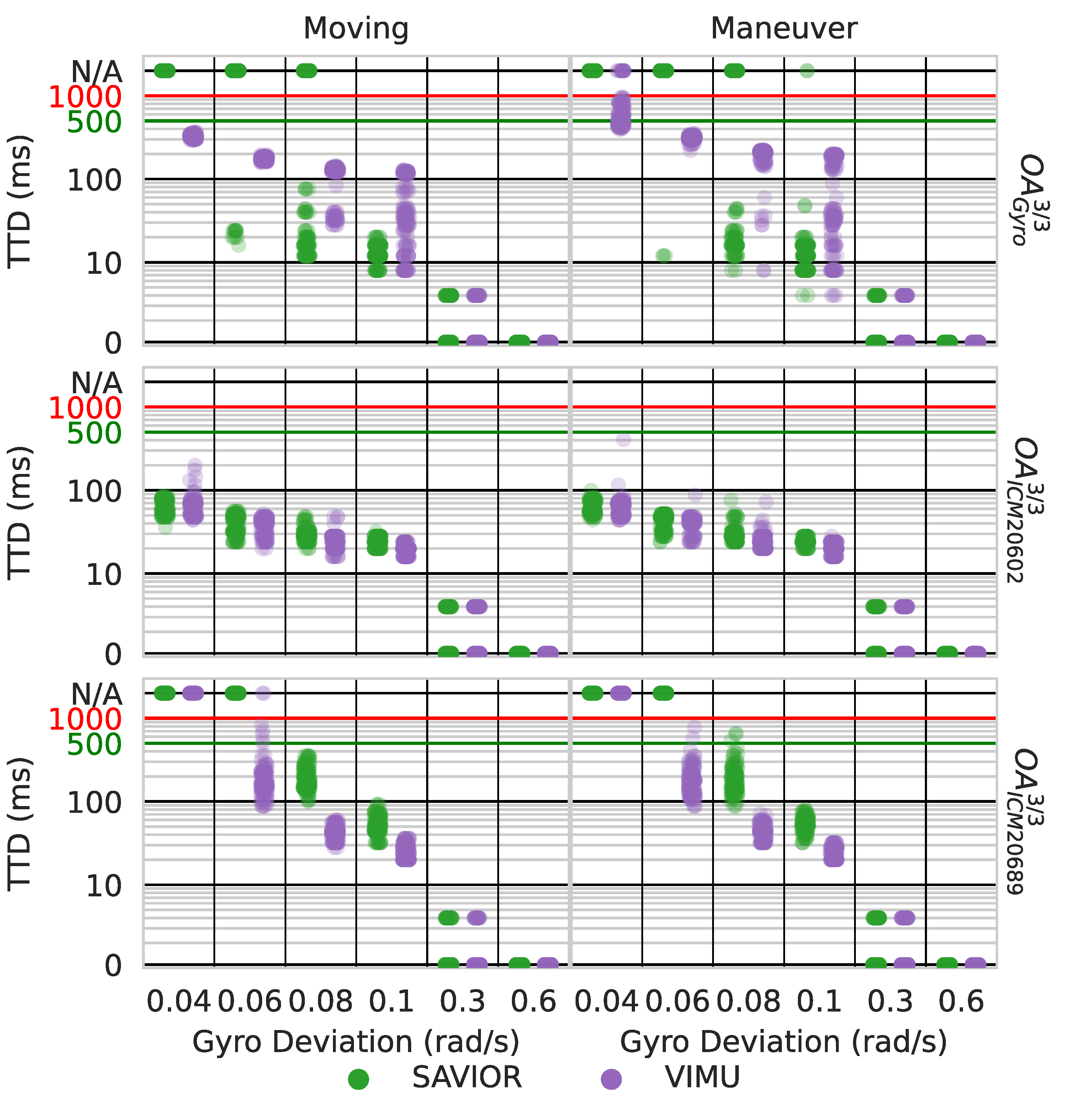}
\vspace{-2ex}
    \caption{TTD on Overt Gyro Attacks in \textit{Maneuver} Mission}
    \label{fig:time_to_detect_maneuver_overt}
\vspace{-2ex}
\end{figure}

\begin{figure}[tb!]
    \centering
    \includegraphics[width=0.85\linewidth]{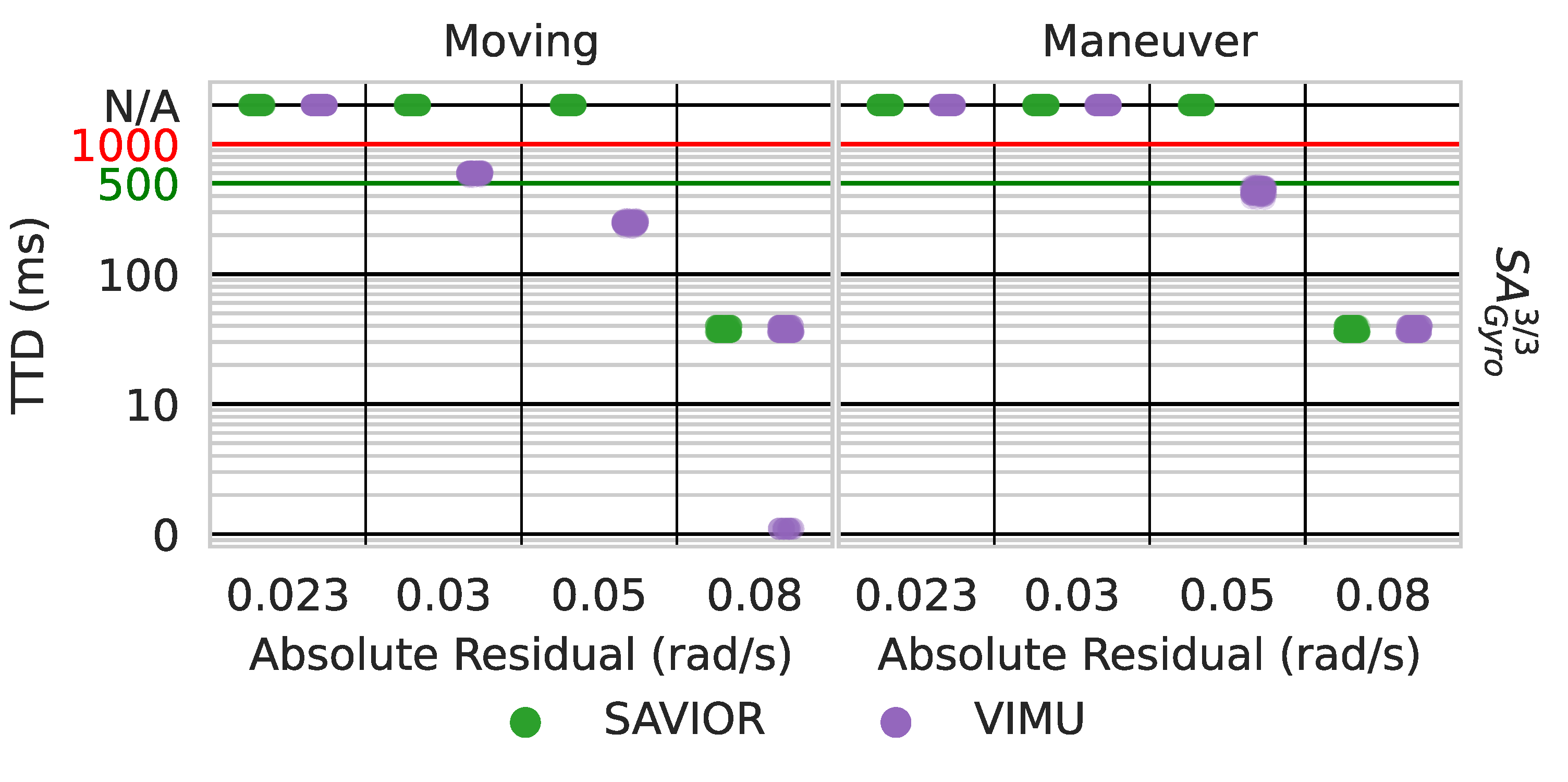}
\vspace{-2ex}
    \caption{TTD on Stealthy Gyro Attack in \textit{Maneuver} Mission}
    \label{fig:time_to_detect_maneuver_stealthy}
\vspace{-2ex}
\end{figure}

\begin{figure}[tb!]
    \centering
    \includegraphics[width=0.9\linewidth]{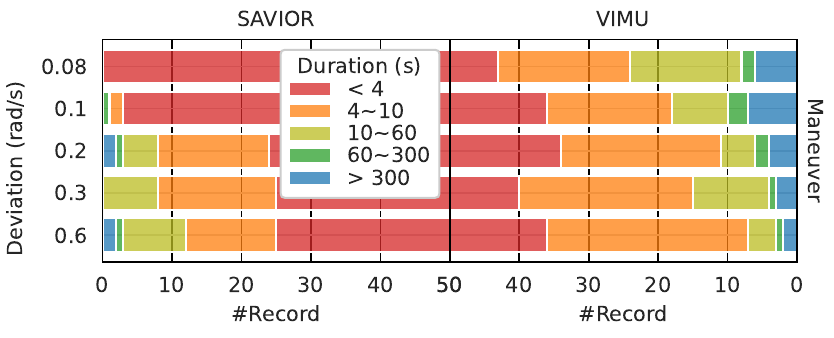}
\vspace{-2ex}
    \caption{Recovery Duration in \textit{Maneuver} Mission}
    \label{fig:recovery_duration_with_savior_maneuver}
\vspace{-2ex}
\end{figure}

\revise{We investigate \sysname{}'s detection effectiveness in the \textit{Maneuver} scenario by comparing it with \savior{} under the gyroscope attacks.
Fig.~\ref{fig:time_to_detect_maneuver_overt} shows the log-scaled TTDs against overt attacks, while Fig.~\ref{fig:time_to_detect_maneuver_stealthy} presents the results against stealthy attacks.
In Fig.~\ref{fig:time_to_detect_maneuver_overt}, both approaches exhibit similar TTDs in the \textit{Maneuver} scenario as in the \textit{Moving} scenario. Due to the relaxed detection threshold $\tau_{ema}$, \sysname{}'s TTD in the \textit{Maneuver} scenario is slightly longer than in the \textit{Moving} scenario. Such threshold relaxation also makes \sysname{} hard to detect the stealthy attack $\SAGyro$(0.03) in the \textit{Maneuver} scenario, as shown in Fig.~\ref{fig:time_to_detect_maneuver_stealthy}. In general, \sysname{}'s advantage in detection effectiveness remains unaffected despite the maneuvers. We also observed similar comparative advantages in recovery duration.
The original \savior{} does not contains a recovery component, so we extends \savior{} with the recovery monitor described in Section~\ref{subsec:method_recovery}.
As presented in Fig.~\ref{fig:recovery_duration_with_savior_maneuver}, even with the maneuvers, \sysname{} outperforms \savior{} in recovery duration, which validates the contribution of our fine-grained physical model and proposed buffer safeguard.
}

\subsection{External Disturbance}\label{app:evaluation-wind}

External disturbances, e.g., wind gusts, disturb the drone in different ways~\cite{tomic2020simultaneous}, and they can raise false alarms that will disrupt the recovery of the system states.
Although the autopilot can compensate on these disturbances, it relies upon a precise state estimate, which validates our usage of the wind velocity term in our aerodynamic drag model ($\Windspeed$ in \eqref{eq:drag_accel}).
The wind speed data required to calculate this term is measurable by an anemometer~\cite{quinonez2020savior} or estimated by the drag acceleration~\cite{leishman2014quadrotors}.
\revise{However, we encounter difficulties in equipping the quadcopter with a wind-speed sensor and integrating it with the autopilot board to measure the wind in flight.}
Therefore, we use simulation to evaluate \sysname{}'s performance in the presence
of wind disturbances.

\begin{figure}[tb!]
    \centering
    \includegraphics[width=0.9\linewidth]{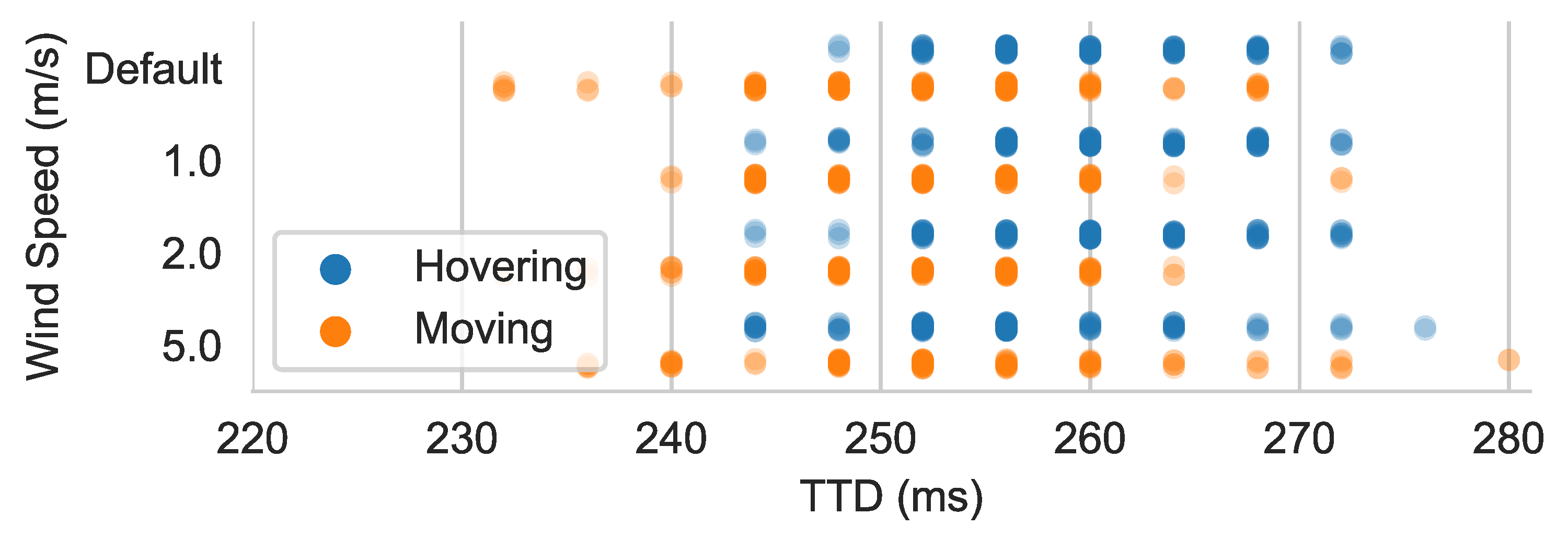}
\vspace{-2ex}
    \caption{TTD on $\SAGyro$ at Different Wind Speeds}
    \label{fig:wind_performance_ttd_duration}
\vspace{-2ex}
\end{figure}

\begin{figure}[tb!]
    \centering
    \includegraphics[width=0.9\linewidth]{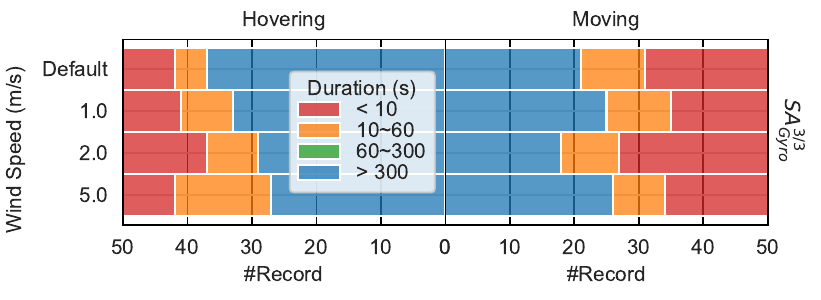}
\vspace{-2ex}
    \caption{Recovery Duration at Different Wind Speeds}
    \label{fig:wind_performance_recovery_duration}
\vspace{-2ex}
\end{figure}

\begin{figure}[tb!]
    \centering
    \includegraphics[width=0.86\linewidth]{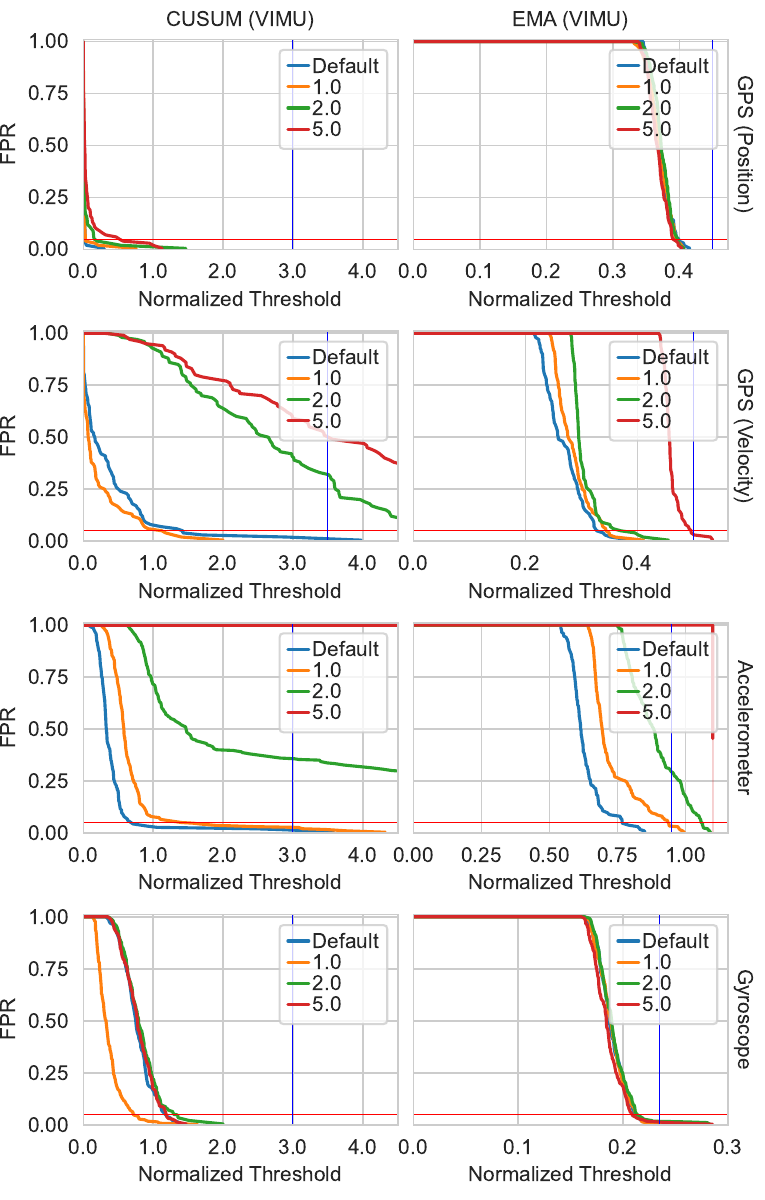}
\vspace{-2ex}
    \caption{FPR-Threshold Curves at Different Wind Speeds (Red Line: 5\% FPR ; Blue Line: Optimal Threshold w.r.t. Optimal Detector Parameters)}
    \label{fig:wind_performance_param_selection}
\vspace{-2ex}
\end{figure}

We configure jMAVSim with different wind speeds and compare the results with the default wind setting.
In jMAVSim, the wind velocity starts from a base vector $\mu$ and follows a random walk process subjected to $\mathcal{N}(0,\,\sigma^{2})$. By default, $\mu = [0.0, 0.0, 0.0]$ m/s and $\sigma = [6.0, 8.0, 0.0]$ m/s in the NED frame. We inject the north-axis wind with $\mu' =$ $[1.0/2.0/5.0, 0.0, 0.0]$ m/s for the wind speeds of 1~m/s, 2~m/s, and 5~m/s, respectively.
Then, we evaluate how the wind disturbance impacts the TTD and the effective recovery duration in stealthy attack $\SAGyro$.
Fig.~\ref{fig:wind_performance_ttd_duration} shows that the TTD in $\SAGyro$ remains steady at different wind speeds.
In Fig.~\ref{fig:wind_performance_recovery_duration}, we only observed a slight decrease in average recovery duration (177.8 s at default setting, 178.0 s at 1.0 m/s, 145.8 s at 2.0 m/s, and 163.4 s at 5.0 m/s). In \textit{Moving} missions, \sysname{} even achieves a longer recovery duration in the higher wind speeds, indicating wind disturbance is not a dominant factor in the recovery of these tasks.

Because adapting different detector parameters to various wind speeds is challenging, we examine the generalizability of the optimal detector parameters obtained at the default wind setting to the wind disturbance environments.
Specifically, we use these parameters to plot the FPR-Threshold curves at different wind speeds (following Appendix~\ref{app:Param_Selection}).
Fig.~\ref{fig:wind_performance_param_selection} presents the FPR-Threshold curves for GPS and IMU.
The FPRs on the gyroscope and GPS position detectors demonstrate that for both CUSUM and EMA detectors, the optimal detector parameters at the default wind setting can tolerate the windy environments, i.e., all curves meet the 5\% FPR requirements specified in the parameter selection (Appendix~\ref{app:Param_Selection}). In contrast,
the FPRs on the accelerometer and GPS velocity detectors show that the false positives increase drastically with wind disturbance.
At 2.0~m/s and 5.0~m/s wind speeds, only the EMA detector for GPS velocity can tolerate the wind.
The detection parameters of two CUSUM detectors and the EMA detector for the accelerometer need to be reconfigured for wind disturbance.

\end{document}